
%
%
\magnification=\magstep 1
\baselineskip=20 pt
\font\bigbf = cmbx10 scaled\magstep 1
 1
 1

\line{}

\noindent
\line{\hfill\hfill\hfill cond-mat/9504063}

\centerline{\bigbf \hfill Quantum Boltzmann equation of \hfill}
\centerline{\bigbf \hfill composite fermions interacting with
a gauge field \hfill}

\vskip 0.3cm

\centerline{\hfill Yong Baek Kim, Patrick A. Lee, and Xiao-Gang Wen \hfill}
\vskip 0.1cm
\centerline{\hfill \it Department of Physics, Massachusetts Institute of
Technology \hfill}
\centerline{\hfill \it Cambridge, Massachusetts 02139 \hfill}

\vskip 0.3cm

\centerline{\hfill April 17, 1995 \hfill}

\vskip 0.5cm

\centerline{\hfill ABSTRACT \hfill}

\vskip 0.3cm

\midinsert
\narrower
{\noindent\tenrm

We derive the quantum Boltzmann equation (QBE) of composite fermions
at/near the $\nu = 1/2$ state using the non-equilibrium
Green's function technique.
The lowest order perturbative correction to the self-energy
due to the strong gauge field fluctuations suggests that
there is no well defined Landau-quasi-particle.
Therefore, we cannot assume the existence of the
Landau-quasi-particles {\it a priori} in the derivation
of the QBE.
Using an alternative formulation, we derive the
QBE for the generalized Fermi surface displacement
which corresponds to the local variation of the chemical
potential in momentum space.
{}From this QBE, one can understand in a
unified fashion the Fermi-liquid behaviors of the density-density and
the current-current correlation
functions at $\nu = 1/2$ (in the long wave length and the low
frequency limits) and the singular behavior of the energy gap
obtained from the finite temperature activation behavior of
the compressibility near $\nu = 1/2$.
Implications of these results to the recent experiments are
also discussed.

}
\endinsert

\vfill\vfill\vfill
\break

\centerline{\bf I. INTRODUCTION}

Since the discovery of the integer (IQH) and
fractional quantum Hall (FQH)
effects the two-dimensional electron system in strong magnetic
fields has often surprised us.
Among recent developements, a lot of attention has been given
to the appearance of the new metallic state at the filling
fraction $\nu = 1/2$ [1] and the associated Shubnikov-de Haas
oscillations of the longitudinal resistance around $\nu = 1/2$ [2,3].
The similarity between these phenamena near $\nu = 1/2$ and
those of electrons in weak magnetic fields was successfully
explained by the composite fermion approach [4].
Using the fermionic Chern-Simons gauge theory of the composite
fermions [5,6], Halperin, Lee, and Read (HLR) developed a theory that
describes the new metallic state at $\nu = 1/2$ [6].

A composite fermion is obtained by attaching an even number $2n$
of flux quanta to an electron and the transformation can be
realized by introducing an appropriate Chern-Simons gauge field [4-6].
At the mean field level, one takes into account only
the average of the statistical
magnetic field due to the attached magnetic flux.
If the interaction between fermions is ignored, the system can be
described as the free fermions in an effective magnetic field
$\Delta B = B - B_{1/2n}$, where $B_{1/2n} = 2 n n_e h c / e$
is the averaged statistical magnetic field and $n_e$ is the
density of electrons.
Therefore, in the mean field theory, the FQH states with
$\nu = {p \over 2np + 1}$ can be described as the IQH state of
the composite fermions with $p$ filled Landau levels occupied
in an effective magnetic field $\Delta B$ [4-6].
In particular, $\Delta B = 0$ at the filling fractions
$\nu = 1/2n$ so that the ground state of the system is the filled
Fermi sea with a well defined Fermi wave vector $k_F$ [6,7].
As a result, the Shubnikov-de Haas oscillations near $\nu = 1/2$
can be explained by the presence of a well defined Fermi wave vector
at $\nu = 1/2$ [6].
The mean field energy gap of the system with
$\nu = {p \over 2p + 1}$ in the $p \rightarrow \infty$ limit is
given by $E_{\rm g} = {e \Delta B \over m c}$, where $m$ is
the mass of the composite fermions.
Note that, in the large $\omega_c$ limit,
the finite $m$ is caused by the Coulomb
interaction between the fermions. The effective mass $m$ should be
chosen such that the Fermi energy $E_F$ is given by the Coulomb energy scale.

There are a number of experiments which show that there is a well
defined Fermi wave vector at $\nu = 1/2$ [8-10].
They observed the geometrical resonances between the semiclassical
orbit of the composite fermions and another length scale
artificially introduced to the system near $\nu =1/2$.

However, it is possible that the fluctuations
and the two-particle interactions, which
are ignored in the mean field theory, are very important.
Note that the density fluctuations correspond to the
fluctuations of the statistical magnetic field.
Therefore, the density fluctuations
above the mean field state induces the gauge field fluctuations [5,6].
If the fermions are interacting via a two-particle interaction
$v ({\bf q}) = V_0 / q^{2 - \eta} (1 \le \eta \le 2)$, the effects
of the gauge field fluctuations can be modified.
In fact, the gauge
field fluctuations become more singular as the interaction
range becomes shorter (larger $\eta$).
The reason is that the longer range interaction (smaller $\eta$)
suppresses more effectively the density fluctuations, thus it
induces the less singular gauge field fluctuations.
Therefore, it is important to examine whether the mean field
Fermi-liquid state is stable against the gauge field fluctuations
which also includes the effects of the two-particle interaction.

One way to study the stability of the mean field Fermi-liquid state
is to examine the low energy behavior of the self-energy correction
induced by the gauge field fluctuations.
It is found that the most singular contribution to the
self-energy $\Sigma ({\bf k}, \omega)$
comes from the transverse part of the gauge field fluctuations [6,11].
The lowest order perturbative correction to the self-energy
(due to the transeverse gauge field) is calculated by
several authors [6,11].
It turns out that ${\rm Re} \ \Sigma \sim {\rm Im} \ \Sigma
\sim \omega^{2 \over 1 + \eta}$ for $1 < \eta \le 2$ and
${\rm Re} \ \Sigma \sim \omega \ {\rm ln} \ \omega$,
${\rm Im} \ \Sigma \sim \omega$ for $\eta = 1$ (Coulomb interaction).
Thus the Landau criterion for the quasi-particle is violated in
the case of $1 < \eta \le 2$ and the case of $\eta = 1$ shows the
marginal Fermi liquid behavior.
In either cases, the effective mass of the fermions diverges, as
$m^* / m \propto |\xi_{k}|^{-{\eta - 1 \over \eta + 1}}$ for
$1 < \eta \le 2$ and as $m^* / m \propto |{\rm ln} \ \xi_k|$
for $\eta = 1$, where $\xi_k = {k^2 \over 2m} - \mu$ and $\mu$
is the chemical potential [6].

In a self-consistent treatment of the self-energy [6], the energy gap
of the system in the presence of a small effective magnetic
field $\Delta B$ can be determined as
$E_{\rm g} \propto |\Delta B|^{1 + \eta \over 2}$
for $1 < \eta \le 2$ and
$E_{\rm g} \propto {|\Delta B| \over |{\rm ln} \ \Delta B|}$
for $\eta = 1$.
Therefore, the energy gap of the system vanishes faster than the
mean field prediction or equivalently the effective mass diverges
in a singular way as $\nu = 1/2$ is approached.
These results suggest that the effective Fermi velocity of the fermion
$v^*_F$ goes to zero at $\nu = 1/2$ even though the Fermi wave vector
$k_F$ is finite and the quasi-particles have a very short life time
$\tau \approx (T / \varepsilon_{F})^{- 2 / (1 + \eta)}
(1 / \varepsilon_{F})$, where $T$ is the temperature and
$\varepsilon_{F}$ is the Fermi energy.
However, the recent magnetic focusing experiment [10] suggests that the
fermion has a long life time or a long mean free path which seems
inconsistent with the above picture.

Since the one-particle Green's function is not gauge-invariant,
the singular self-energy could be an artifact of the gauge choice.
To address this question, recently we examined the lowest order
perturbative corrections to the
gauge-invariant density-density and the current-current correlation
functions [12]. It is found that there are important cancellations
between the self-energy corrections and the vertex corrections
due to the Ward-identity [12,13].
As a result, the density-density and the current-current correlation
functions show a Fermi-liquid behavior for all ratios of $\omega$
and $v_F q$ [12].
In particular, the edge of the particle-hole continuum $\omega = v_F q$
is essentially not changed, which may suggest a finite effective mass.
{}From the current-current correlation function, the transport scattering
rate (due to the transverse part of the gauge field) is given by
$1 / \tau_{\rm tr} \propto \omega^{4 \over 1 + \eta} \ll \omega$
after the cancellation
(The scattering rate would be much larger
$1 / \tau_{\rm tr} \propto \omega^{2 \over 1 + \eta} \gg \omega$
had we ignored the vertex correction) [12].
Therefore, the fermions have a long transport life time which
explains a long free path in the magnetic focusing experiment.
{}From these results, one may suspect whether the divergent mass obtained
from the self-energy has any physical meaning.

However, due to the absence of the underlying quasi-particle picture,
we cannot simply conclude that the
fermions have a finite effective mass associated with the long life time
which was obtained from the small $q$ and $\omega$ behaviors of the
density-density and the current-current correlation functions.
In fact, it is found that $2 k_F$ response functions show singular
behaviors compared to the usual Fermi liquid theory [13].
We also like to mention that
the recent experiments on the Shbunikov-de Hass oscillations [3]
have observed some features which were interpretated as a sign of
the divergent
effective mass of the fermions as $\nu = 1/2$
is approached.
The experimentally determined effective mass diverges in a more singular
way than any theoretical prediction.
However, their determination of the effective mass is based on a theory
for the non-interacting fermions and also the disorder effect is very
important near $\nu = 1/2$ because the static fluctuations of the
density due to the impurities induces an additional static random
magnetic field.
Since there is no satisfactory theory in the presence of disorder,
it is difficult to compare the present theory and the experiments.

In order to answer the question about the effective mass,
it is important to examine other gauge-invariant quantities
which may potentially show a divergent effective mass.
In a recent paper [14], we calculated the lowest order perturbative
correction to the compressibility with a fixed $\Delta B$, which
shows a thermally activated behavior when the chemical potential
lies exactly at the middle of the successive effective Landau levels.
It turns out that the corrections to the activation energy gap and
the corresponding effective mass are singular and consistent with
the previous self-consistent treatment of the self-energy [6].
Thus it is necessary to understand the apparently different behaviors
of the density-density correlation function at $\nu = 1/2$ and
the activation energy gap determined from the compressibility
near $\nu =1/2$.

One resolution of the problem was suggested by Stern and Halperin [15]
within the usual Landau-Fermi-liquid theory framework.
The idea is that both of the effective mass and the
Landau-interaction-function are singular in such a way
that they cancel
each other in the density-density correlation function.
Recently, Stern and Halperin [15] put forward this idea and construct
a Fermi-liquid-theory of the fermion-gauge system
in the case of Coulomb interaction.
Even though the use of the Landau-Fermi-liquid theory or equivalently
the existence of the well defined quasi-particles can be
{\it marginally} justified in the case of the Coulomb interacion,
we feel that it is necessary to construct a more general framework
which applies to the arbitrary two-particle interaction
($1 < \eta \le 2$ as well as $\eta = 1$) and
allows us check the validity of the Fermi liquid theory and
to judge when the divergent mass shows up.
In particular, it is worthwhile to provide a unified picture for
understanding the previous theoretical studies [16-24].

In the usual Fermi-liquid theory, the QBE of the quasi-particles
provides the useful informations about the low lying excitations
of the system.
Our objective is to construct a similar QBE which describes all the
low energy physics of the composite fermion system.
One important difficulty we are facing here
is that we cannot assume the existence
of the quiasi-particles {\it a priori} in the derivation
of the QBE even though the conventional derivation of the QBE
of the Fermi-liquid theory relies on the existence of these
quasi-particles.
Following closely the work of Prange and Kadanoff [25] about the
electron-phonon system, where there is also no well defined
quasi-particle at temperatures high compared with the Debye
temperature,
we concentrate on a generalized Fermi surface
displacement which, in our case, corresponds to the local
variation of the chemical potential in momentum space.
Due to the non-existence of a well defined quasiparticle, the usual
distribution function $n_k$ in the momentum space cannot be
described by a closed equation of motion.
However we will see later that the generalized Fermi surface
displacement does satisfy a closed equation of motion.
This equation of motion will be also called as QBE.

We use the non-equilibrium Green's function
technique [26-28] to derive the new QBE and calculate the generalized
Landau-interaction-function which has the frequency dependence
as well as the usual angular dependence due to the retarded
nature of the gauge interaction.
The QBE at $\nu = 1/2$ consists of three parts. One is the contribution
from the self-energy correction which gives the singular
mass correction, the other one comes from the generalized
Landau-interaction-function, and finally it contains
the collision integral.
These quantities are calculated to the lowest order in the coupling
to the gauge field.

By studing the dynamical properties of the collective modes
using the QBE,
we find that the smooth fluctuations of the Fermi surface
(or the small angular momentum modes) show the usual Fermi-liquid
behavior, while the rough fluctuations (or the large angular
momentum modes) show the singular behavior determined by
the singular self-energy correction. Here the angular momentum
is the conjugate variable of the angle measured from a given direction
in momentum space.
There is a forward scattering cancellation
between the singular self-energy correction and the singular
(generalized) Landau-interaction-function and a similar
cancellation exists in the collision integral as far as
the small angular momentum modes $l < l_c$
($l_c \propto \Omega^{-{1 \over 1 + \eta}}$, where $\Omega$ is the
small external frequency) are concerned.
However, in the case of the large angular momentum modes $l > l_c$,
the contribution from the Landau-interaction-function
becomes very small so that the self-energy correction
dominates and the collision integral also cannot be
ignored in general.
In this case the behaviors of the low lying modes are
very different from those in the Fermi liquids.

If we ignore the collision integral, it can be shown that
the system has a lot of collective modes between
$\Omega \propto q^{1 + \eta \over 2}$ ($1 < \eta \le 2$),
$\Omega \propto q / |{\rm ln} \ q|$ ($\eta = 1$) and
$\Omega = v_F q$ while there is the particle-hole continuum
below $\Omega \propto q^{1 + \eta \over 2}$ ($1 < \eta \le 2$),
$\Omega \propto q / |{\rm ln} \ q|$ ($\eta = 1$).
The distinction between these two types of low lying excitations
are obscured by the existence of the collision integral.

{}From the above results, we see
that the density-density and the current-current
correlation functions, being  dominated by the small angular momentum
modes $l < l_c$, show the usual Fermi-liquid behavior.
On the other hand, the energy gap away from $\nu=1/2$
is determined by the behaviors
of the large angular momentum modes $l > l_c$ so that the singular
mass correction shows up in the energy gap of the system.

The outline of the paper is as follows.
In section II, we introduce the model and explain the way we
contruct the QBE without assuming the existence of the
quasi-particles.
In section III, the QBE for the generalized distribution function
is derived for $\Delta B = 0$.
In section IV, we construct the QBE for the generalized Fermi surface
displacement for $\Delta B = 0$.
We also determine the generalized Landau-interaction-function
and discuss its consequences.
In section V, The QBE in the presence of a small $\Delta B$
is constructed and the energy gap of the system is determined.
In section VI, We discuss the collective excitations of the
system for the cases of $\Delta B = 0$ and $\Delta B \not= 0$.
We conclude the paper and discuss the implications of our
results to experiments in section VII.
We concentrate on the zero temperature case in the main text
and provide the derivation of the QBE at finite temperatures
in the appendix,
which requires some special treatments compared to the zero
temperature counterpart.

\vskip 0.5cm

\centerline{\bf II. THE MODEL AND THE QUANTUM BOLTZMANN EQUATION}
\centerline{\bf IN THE ABSENCE OF THE QUASI-PARTICLES}

The two dimensional electrons interacting via a two-particle
interaction can be transformed to the composite fermions interacting
via the same two-partice interaction
and also coupled to an appropriate Chern-Simons
gauge field which appears due to the statistical
magnetic flux quanta attached to each electron [5,6].
The model can be constructed as follows $(\hbar = e = c = 1$).
$$
Z = \int \ D\psi \ D\psi^*  \ Da_{\mu} \ e^{\ i \int dt \ d^2 r \ {\cal L}} \ ,
\eqno{(1)}
$$
where the Lagrangian density ${\cal L}$ is
$$
\eqalign{
{\cal L} &= \psi^* (\partial_{0} + ia_0 - \mu) \psi - {1 \over 2m} \psi^*
(\partial_i - i a_i + i A_i)^2 \psi \cr
&- {i \over 2 \pi {\tilde \phi}} \ a_0 \varepsilon^{ij} \partial_i a_j +
{1 \over 2} \int d^2 r' \ \psi^* ({\bf r}) \psi ({\bf r}) \
v ({\bf r} - {\bf r'}) \ \psi^* ({\bf r'}) \psi ({\bf r'}) \ ,
}
\eqno{(2)}
$$
where $\psi$ represents the fermion field and ${\tilde \phi}$ is an even
number $2n$ which is the number of flux quanta attached to an electron,
and $v ({\bf r}) \propto V_0 / r^{\eta}$ is the Fourier transform of
$v ({\bf q}) = V_0 / q^{2 - \eta} \ (1 \le \eta \le 2)$ which represents
the interaction between the fermions.
${\bf A}$ is the external vector potential ($B = \nabla \times {\bf A}$)
and we choose the Coulomb gauge $\nabla \cdot {\bf a} = 0$ for
the Chern-Simons gauge field.
Note that the integration over $a_0$ enforces the following constraint:
$$
\nabla \times {\bf a} = 2 \pi {\tilde \phi} \
\psi^* ({\bf r}) \psi ({\bf r}) \ ,
\eqno{(3)}
$$
which represents the fact that ${\tilde \phi}$ number of flux quanta
are attached to each electron.

The saddle point of the action is given by the following conditions:
$$
\nabla \times \langle {\bf a} \rangle = 2 \pi {\tilde \phi} \ n_e = B_{1/2n}
\ \ {\rm and} \ \  \langle a_0 \rangle = 0 \ .
\eqno{(4)}
$$
Therefore, at the mean field level, the fermions see an
effective magnetic field ($\Delta {\bf A} = {\bf A} - \langle {\bf a}
\rangle$):
$$
\Delta B = \nabla \times \Delta {\bf A} = B - B_{1/2n} \ ,
\eqno{(5)}
$$
which becomes zero at the Landau level filling factor $\nu = 1/2n$.
The IQH effect of the fermions may appear when the effective
Landau level filling factor $p = {2 \pi n_e \over \Delta B}$ becomes
an integer.
This implies that the real external magnetic field is given by
$B = B_{1/2n} + \Delta B = 2 \pi n_e \ \left (
{2np + 1 \over p} \right )$ which corresponds to a FQH state
of electrons with the filling factor $\nu = {p \over 2np +1}$.

The fluctuations of the Chern-Simons gauge field,
$\delta a_{\mu} = a_{\mu} - \langle a_{\mu} \rangle$, can be
incorperated as follows.
$$
Z = \int \ D\psi \ D\psi^*  \ D \delta a_{\mu} \
e^{\ i \int dt \ d^2 r \ {\cal L}} \ ,
\eqno{(6)}
$$
where
$$
\eqalign{
{\cal L} &= \psi^* (\partial_{0} + i \ \delta a_0 - \mu) \psi -
{1 \over 2m} \psi^* (\partial_i - i \ \delta a_i + i \ \Delta A_i)^2 \psi \cr
&- {i \over 2 \pi {\tilde \phi}} \ \delta a_0 \ \varepsilon^{ij} \
\partial_i \ \delta a_j +
{1 \over 2} \int d^2 r' \ (\nabla \times \delta {\bf a} ({\bf r})) \
v ({\bf r} - {\bf r'}) \ (\nabla \times \delta {\bf a} ({\bf r'})) \ ,
}
\eqno{(7)}
$$
After integrating out the fermions and including gauge field fluctuations
within the random phase approximation (RPA) [6], the effective action of
the gauge field can be obtained as
$$
S_{\rm eff} = {1 \over 2} \int {d^2 q \over (2 \pi)^2} \
{d\omega \over 2 \pi}
\ \delta a^*_{\mu} ({\bf q}, \omega) \
D^{-1}_{\mu \nu} ({\bf q},\omega,\Delta B)
\ \delta a_{\nu} ({\bf q}, \omega) \ ,
\eqno{(8)}
$$
where $D^{-1}_{\mu \nu} ({\bf q},\omega,\Delta B)$ was calculated
by several authors [6,29,30].
For our purpose, the $2 \times 2$ matrix form for $D^{-1}_{\mu \nu}$
is sufficient so that $\mu, \nu = 0, 1$ and $1$ represents the
direction that is perpendicular to ${\bf q}$.
In particular, when $\Delta B = 0$, the gauge field propagator
has the following form [6]:
$$
D^{-1}_{\mu \nu} ({\bf q},\omega) =
\pmatrix{ {m \over 2 \pi} & -i {q \over 2 \pi {\tilde \phi}} \cr
i {q \over 2 \pi {\tilde \phi}} & -i \gamma {\omega \over q} +
{\tilde \chi} (q) q^2} \ ,
\eqno{(9)}
$$
where $\gamma = {2 n_e \over k_F}$ and ${\tilde \chi} (q) =
{1 \over 24 \pi m} + {v (q) \over (2 \pi {\tilde \phi})^2}$.
Since the most singular contribution to the self-energy correction
comes from the transverse part of the gauge field [6,11], we concentrate
on the effect of the transverse gauge field fluctuations.
In the infrared limit, the transeverse gauge field propagator
can be taken as [12,14]
$$
D_{11} ({\bf q},\omega) =
{1 \over -i \gamma {\omega \over q} + \chi q^{\eta}} \ ,
\eqno{(10)}
$$
where $\chi = {1 \over 24 \pi m} +
{V_0 \over (2 \pi {\tilde \phi})^2}$ for $\eta = 2$ and
$\chi = {V_0 \over (2 \pi {\tilde \phi})^2}$ for $1 \le \eta < 2$.

Before explaining the way we construct the QBE for the fermion-gauge-field
system in which there is no well defined Landau-quasi-particle in general,
we review the usual derivation of the QBE for the Fermi-liquid with
well defined quasi-particles [19,21].
The QBE is nothing but the equation of motion of the fermion distribution
function. Therefore, it can be derived from the equation of motion
of the non-equilibrium one-particle Green's function.
Following Kadanoff and Baym [26], let us consider the following one-particle
Green's function.
$$
G^< (x_1, x_2) = i \langle \psi^{\dagger} (x_2) \psi (x_1) \rangle \ ,
\eqno{(11)}
$$
where $x_1 = ({\bf r}_1, t_1)$ and $x_2 = ({\bf r}_2, t_2)$.
At non-equilibrium, $G^< (x_1, x_2)$ does not satisfy the translational
invariance in space-time so that it cannot be written as $G^< (x_1 - x_2)$.
By the following change of variables
$$
\eqalign{
({\bf r}_{\rm rel}, t_{\rm rel}) = x_1 - x_2 \ \ {\rm and} \ \
({\bf r}, t) = (x_1 + x_2) / 2 \ ,}
\eqno{(12)}
$$
$G^< (x_1, x_2)$ can be written as
$$
G^< ({\bf r}_{\rm rel}, t_{\rm rel}; {\bf r}, t)
= i \langle \psi^{\dagger} ({\bf r} - {{\bf r}_{\rm rel} \over 2},
t - {t_{\rm rel} \over 2}) \
\psi ({\bf r} + {{\bf r}_{\rm rel} \over 2}, t + {t_{\rm rel} \over 2})
\rangle \ .
\eqno{(13)}
$$
By the Fourier transformation for the relative coordinates $t_{\rm rel}$
and ${\bf r}_{\rm rel}$, we get $G^< ({\bf p}, \omega; {\bf r}, t)$.
At equilibrium, $G^<$ can be written as [26-28]
$$
G^<_0 ({\bf p}, \omega) = i f_0 (\omega) A ({\bf p}, \omega) \ ,
\eqno{(14)}
$$
where $f_0 (\omega) = 1 / (e^{\omega / T} + 1)$ is the equilibrium
Fermi distribution function and ($\Sigma^R$ is the retarded
self-energy)
$$
A ({\bf p}, \omega) = {-2 \ {\rm Im} \ \Sigma^R \ ({\bf p}, \omega)
\over (\omega - \xi_p - {\rm Re} \ \Sigma^R ({\bf p}, \omega))^2
+ ({\rm Im} \ \Sigma^R ({\bf p}, \omega))^2} \ .
\eqno{(15)}
$$

In the usual Fermi-liquid theory, ${\rm Im} \ \Sigma^R \ll \omega$
so that $A ({\bf p}, \omega)$ is a peaked function of $\omega$ around
$\omega = \xi_p + {\rm Re} \ \Sigma^R$.
In this case, the equilibrium spectral function can be taken as [26-28]
$$
A ({\bf p}, \omega) = 2 \pi
\delta (\omega - \xi_p - {\rm Re} \ \Sigma^R ({\bf p}, \omega)) \ .
\eqno{(16)}
$$
Using this property, if the system is not far away from the equilibrium,
one can construct a closed equation for the fermion distribution function
$f ({\bf p}, {\bf r}, t)$ [26-28], which is the QBE.
The linearized QBE of $\delta f ({\bf p}, {\bf r}, t) =
f ({\bf p}, {\bf r}, t) - f_0 ({\bf p})$, where $f_0 ({\bf p})$
is the equilibrium distribution function, is the QBE of the
quasi-particles in the Fermi-liquid theory.
{}From this QBE, the equation of motion for the Fermi surface
deformation, which is defined as [26-28]
$$
\nu (\theta, {\bf r}, t) = \int d |{\bf p}| \
\delta f ({\bf p}, {\rm r}, t) \ ,
\eqno{(17)}
$$
can be also constructed.

In the case of the fermion-gauge-field system, as mentioned in the
introduction, ${\rm Im} \ \Sigma^R (\omega)$ is larger than $\omega$
($1 < \eta \le 2$) or comparable to $\omega$ ($\eta = 1$), {\it i.e.},
strictly speaking, there is no well defined Landau-quasi-particle
from the viewpoint of perturbation theory.
However, Stern and Halperin [15] showed that, within a self-consistent
treatment, the Fermi-liquid theory can be barely applied to
the case of Coulomb interaction in the sense that
${\rm Re} \ \Sigma^R$ is logarithmically larger than
${\rm Im} \ \Sigma^R$.
Note that, in general,
$A ({\bf p}, \omega)$ at equilibrium is not a peaked function
of $\omega$ anymore in the fermion-gauge-field system.
Because of this, $f ({\bf p}, {\bf r}, t)$ does not satisfy a
closed equation of motion even near the equilibrium.
However, if $\Sigma^R$ is only a function of $\omega$,
$A ({\bf p}, \omega)$ is still a well peaked function of $\xi_p$ around
$\xi_p = 0$ for sufficiently small $\omega$ [25].
This observation leads us to define the following generalized
distribution function [25]
$$
f (\theta, \omega; {\bf r}, t) = -i \int {d \xi_p \over 2 \pi} \
G^< ({\bf p}, \omega; {\bf r}, t) \ ,
\eqno{(18)}
$$
where $\theta$ is the angle between ${\bf p}$ and a given direction.
The linearized quantum Boltzmann equation for
$\delta f (\theta, \omega; {\bf r}, t) = f (\theta, \omega; {\bf r}, t)
- f_0 (\omega)$ can be derived, which is analogous to the QBE of
the quasi-particles in the usual Fermi-liquid theory.
{}From this QBE, one can also construct the equation of motion
for the generalized Fermi surface displacement [25]
$$
u (\theta, {\bf r}, t) = \int {d \omega \over 2 \pi} \
\delta f (\theta, \omega; {\bf r}, t)
\eqno{(19)}
$$
which corresponds to the variation of the local chemical potential
in the momentum space.
This object can be still well defined even in the absence of the sharp
Fermi surface.
This is because one can always define a chemical potential in each
angle $\theta$, which is the energy required to put an additional
fermion in the direction labeled by $\theta$ in the momentum space.
In the next section, we derive the linearized QBE for the generalized
distribution function $\delta f (\theta, \omega; {\bf r}, t)$.

\vskip 0.5cm

\centerline{\bf III. QUANTUM BOLTZMANN EQUATION}
\centerline{\bf FOR THE GENERALIZED DISTRIBUTION FUNCTION}

In the non-equilibrium Green's function formulation, the
following matrices of the Green's function and the self-energy
satisfy the Dyson's equation [28]
$$
{\tilde G} =
\pmatrix{G_t & - G^< \cr
G^> & - G_{\bar t}} \ \ {\rm and} \ \
{\tilde \Sigma} =
\pmatrix{\Sigma_t & - \Sigma^< \cr
\Sigma^> & - \Sigma_{\bar t}},
\eqno{(20)}
$$
where
$$
\eqalign{
G^> (x_1, x_2) &= - i \langle \psi (x_1) \psi^{\dagger} (x_2) \rangle, \cr
G^< (x_1, x_2) &= i \langle \psi^{\dagger} (x_1) \psi (x_2) \rangle, \cr
G_t (x_1, x_2) &= \Theta (t_1 - t_2) G^> (x_1, x_2) +
		               \Theta (t_2 - t_1) G^< (x_1, x_2), \cr
G_{\bar t} (x_1, x_2) &= \Theta (t_2 - t_1) G^> (x_1, x_2) +
		                      \Theta (t_1 - t_2) G^< (x_1, x_2),
}
\eqno{(21)}
$$
and $\Sigma^>, \Sigma^<, \Sigma_t, \Sigma_{\bar t}$ are the associated
self-energies. $\Theta (t) = 1$ for $t > 0$ and zero for $t < 0$.
$G^R$ (retarded) and $G^A$ (advanced) Green's functions can be
expressed in terms of $G_t$ (time-ordered), $G_{\bar t}$
(antitime-ordered), $G^<$, $G^>$ as follows.
$$
\eqalign{
G^R &= G_t - G^< = G^> - G_{\bar t} \ , \cr
G^A &= G_t - G^> = G^< - G_{\bar t} \ .
}
\eqno{(22)}
$$
Similarly, $\Sigma^R$ and $\Sigma^A$ are given by
$$
\eqalign{
\Sigma^R &= \Sigma_t - \Sigma^< = \Sigma^> - \Sigma_{\bar t} \ , \cr
\Sigma^A &= \Sigma_t - \Sigma^> = \Sigma^< - \Sigma_{\bar t} \ .
}
\eqno{(23)}
$$
The matrix Green's function satisfies the following equations of
motion
$$
\eqalign{
\left [ i {\partial \over \partial t_1} - H_0 ({\bf r}_1) \right ]
{\tilde G} (x_1, x_2) &=
\delta (x_1 - x_2) {\tilde I} + \int d x_3 \ {\tilde \Sigma} (x_1, x_3)
{\tilde G} (x_3, x_2), \cr
\left [ - i {\partial \over \partial t_2} - H_0 ({\bf r}_2) \right ]
{\tilde G} (x_1, x_2) &=
\delta (x_1 - x_2) {\tilde I} + \int d x_3 \ {\tilde G} (x_1, x_3)
{\tilde \Sigma} (x_3, x_2) \ ,
}
\eqno{(24)}
$$
where
$$
H_0 ({\bf r}_1) = - {1 \over 2m} \left (
{\partial \over \partial {\bf r}_1} \right )^2 - \mu
\ \ {\rm and} \ \
H_0 ({\bf r}_2) = - {1 \over 2m} \left (
{\partial \over \partial {\bf r}_2} \right )^2 - \mu \ .
\eqno{(25)}
$$
For our purpose, we need only the equation of motion for $G^<$
$$
\eqalign{
\left [ i {\partial \over \partial t_1} - H_0 ({\bf r}_1) \right ]
G^< (x_1, x_2)
&= \int d x_3 \left [ \Sigma_t (x_1, x_3) G^< (x_3, x_2)
- \Sigma^< (x_1, x_3) G_{\bar t} (x_3, x_2) \right ] \ , \cr
\left [ - i {\partial \over \partial t_2} - H_0 ({\bf r}_2)
\right ] G^< (x_1, x_2)
&= \int d x_3 \left [ G_t (x_1, x_3) \Sigma^< (x_3, x_2)
- G^< (x_1, x_3) \Sigma_{\bar t} (x_3, x_2) \right ] \ .
}
\eqno{(26)}
$$
Taking the difference of the two equations of Eq.(26), and
using the following relations
$$
\eqalign{
G_t &= {\rm Re} \ G^R + {1 \over 2} (G^< + G^>) \ , \cr
G_{\bar t} &= {1 \over 2} (G^< + G^>) - {\rm Re} \ G^R \ ,
}
\eqno{(27)}
$$
we get
$$
\eqalign{
&\left [ \ i {\partial \over \partial t_1} +
i {\partial \over \partial t_2} + {1 \over 2m} \left ( {\partial \over
\partial {\bf r}_1} \right )^2 - {1 \over 2m} \left ( {\partial \over
\partial {\bf r}_2} \right )^2 \ \right ] G^< (x_1, x_2) \cr
&=
\int d x_3 \ \bigl [ \ {\rm Re} \ \Sigma^R (x_1, x_3) G^< (x_3, x_2)
+ \Sigma^< (x_1, x_3) {\rm Re} \ G^R (x_3, x_2) \cr
&\hskip 1.35cm -{\rm Re} \ G^R (x_1, x_3) \Sigma^< (x_3, x_2)
- G^< (x_1, x_3) {\rm Re} \ \Sigma^R (x_3, x_2) \cr
&\hskip 1.35cm + {1 \over 2} \Sigma^> (x_1, x_3) G^< (x_3, x_2)
- {1 \over 2} \Sigma^< (x_1, x_3) G^> (x_3, x_2) \cr
&\hskip 1.35cm - {1 \over 2} G^> (x_1, x_3) \Sigma^< (x_3, x_2)
+ {1 \over 2} G^< (x_1, x_3) \Sigma^> (x_3, x_2) \ \bigr ] \ .
}
\eqno{(28)}
$$

Near equilibrium, one can linearize this equation assuming that
$\delta {\tilde G} = {\tilde G} - {\tilde G}_0$ and
$\delta {\tilde \Sigma} = {\tilde \Sigma} - {\tilde \Sigma}_0$
are small, where ${\tilde G}_0$ and ${\tilde \Sigma}_0$ are
matrices of the equilibrium Green's function and the self-energy.
The Fourier transform ${\tilde G} (p_1, p_2)$
($p_1 = ({\bf p}_1, \omega_1)$, $p_2 = ({\bf p}_2, \omega_2)$)
of ${\tilde G} (x_1, x_2)$ can be written in terms of the new
variables defined by
$$
p = ({\bf p}, \omega) = (p_1 - p_2) / 2 \ \ {\rm and} \ \
q = ({\bf q}, \Omega) = p_1 + p_2 \ .
\eqno{(29)}
$$
Using these variables, the Fourier transformed linearized
equation of $\delta G^< (p, q)$ can be written as
$$
\eqalign{
&[ \ \Omega - v_F |{\bf q}| \ {\rm cos} \ \theta_{\bf pq} \ ] \
\delta G^< (p,q) \cr
&- [ \ {\rm Re} \ \Sigma^R_0 (p + q / 2)
- {\rm Re} \ \Sigma^R_0 (p - q / 2) \ ] \ \delta G^< (p,q) \cr
&+ [ \ G^<_0 (p + q / 2)
- G^<_0 (p - q / 2) \ ] \ \delta ({\rm Re} \ \Sigma^R (p,q)) \cr
&- [ \ \Sigma^<_0 (p + q / 2)
- \Sigma^<_0 (p - q / 2) \ ] \ \delta ({\rm Re} \ G^R (p,q)) \cr
&+ [ \ {\rm Re} \ G^R_0 (p + q / 2)
- {\rm Re} \ G^R_0 (p - q / 2) \ ] \ \delta \Sigma^< (p,q) \cr
&=
G^<_0 (p) \ \delta \Sigma^> (p,q) + \Sigma^>_0 (p) \ \delta G^< (p,q)
- G^>_0 (p) \ \delta \Sigma^< (p,q) - \Sigma^<_0 (p) \ \delta G^> (p,q) \ ,
}
\eqno{(30)}
$$
where $\theta_{\bf pq}$ is the angle between ${\bf p}$ and ${\bf q}$.
In the presence of an external potential $U (q)$, one should add
a term $U (q) \ [ \ G^<_0 (p + q / 2)
- G^<_0 (p - q / 2) \ ]$ in the left hand side of Eq.(30).

We next check that this expression is equivalent to the usual
QBE for $\delta G^< ({\bf p}, \omega; {\bf r}, t)$, where
${\bf r}$ and $t$ are conjugate variables of ${\bf q}$ and $\Omega$.
Note that
$$
F (p + q / 2) - F (p - q / 2) \approx
{\bf q} \cdot {\partial F \over \partial {\bf p}} +
\Omega {\partial F \over \partial \omega} \ ,
\eqno{(31)}
$$
for small $|{\bf q}|$ and $\Omega$.
{}From Eq.(30) and Eq.(31), one can check that
$\delta G^< ({\bf p}, \omega; {\bf r}, t)$, which is the Fourier
transform of $\delta G^< (p,q)$, satisfies the following
equation.
$$
\eqalign{
&[ \ \omega - p^2 / 2m,
\delta G^< ({\bf p}, \omega; {\bf r}, t) \ ] \cr
&- [ \ {\rm Re} \ \Sigma^R_0 ({\bf p}, \omega),
G^< ({\bf p}, \omega; {\bf r}, t) \ ]
- [ \ \delta ({\rm Re} \ \Sigma^R ({\bf p}, \omega)),
G^<_0 ({\bf p}, \omega) \ ] \cr
&- [ \ \Sigma^<_0 ({\bf p}, \omega),
\delta ({\rm Re} \ G^R ({\bf p}, \omega; {\bf r}, t) \ ]
- [ \ \delta \Sigma^< ({\bf p}, \omega; {\bf r}, t),
{\rm Re} \ G^R_0 ({\bf p}, \omega) \ ] \cr
&=
G^<_0 ({\bf p}, \omega) \ \delta \Sigma^> ({\bf p}, \omega; {\bf r}, t)
+ \Sigma^>_0 ({\bf p}, \omega) \ \delta G^< ({\bf p}, \omega; {\bf r}, t) \cr
&- G^>_0 ({\bf p}, \omega) \ \delta \Sigma^< ({\bf p}, \omega; {\bf r}, t)
- \Sigma^<_0 ({\bf p}, \omega) \ \delta G^> ({\bf p}, \omega; {\bf r}, t) \ ,
}
\eqno{(32)}
$$
where $[X, Y]$ is the Poisson braket
$$
[X, Y] = {\partial X \over \partial \omega}
{\partial Y \over \partial t} - {\partial X \over \partial t}
{\partial Y \over \partial \omega} -
{\partial X \over \partial {\bf p}} \cdot
{\partial Y \over \partial {\bf r}} +
{\partial X \over \partial {\bf r}} \cdot
{\partial Y \over \partial {\bf p}} \ .
\eqno{(33)}
$$
Note that this equation is just the linearized version
of the usual QBE for $G^< ({\bf p}, \omega; {\bf r}, t)$ given by [25-28]
$$
\eqalign{
&[ \ \omega - p^2 / 2m -
{\rm Re} \ \Sigma^R ({\bf p}, \omega; {\bf r}, t),
G^< ({\bf p}, \omega; {\bf r}, t) \ ]
- [ \ \Sigma^< ({\bf p}, \omega; {\bf r}, t),
{\rm Re} \ G^R ({\bf p}, \omega; {\bf r}, t) \ ] \cr
&= \Sigma^> ({\bf p}, \omega; {\bf r}, t) G^< ({\bf p}, \omega; {\bf r}, t)
- G^> ({\bf p}, \omega; {\bf r}, t) \Sigma^< ({\bf p}, \omega; {\bf r}, t) \ .
}
\eqno{(34)}
$$

We directly deal with Eq.(30) in momentum space $({\bf q}, \Omega)$
rather than the long time, long wave length expansion in real space
$({\bf r}, t)$ given by Eq.(32).
For simplicity, we assume that the gauge field is in equilibrium.
The non-equilibrium one-loop self-energy correction, which is given by
the diagram in Fig.1, can be written as [27,28]
$$
\eqalign{
\Sigma^< ({\bf p}, \omega) &= \sum_{\bf q} \int^{\infty}_{0}
{d \nu \over \pi} \left | {{\bf p} \times {\hat {\bf q}} \over m}
\right |^2 {\rm Im} \ D_{11} ({\bf q}, \nu) \cr
&\times [ \ (n_0 (\nu) + 1) G^< ({\bf p} + {\bf q}, \omega + \nu)
+ n_0 (\nu) G^< ({\bf p} + {\bf q}, \omega - \nu) \ ] \ , \cr
\Sigma^> ({\bf p}, \omega) &= \sum_{\bf q} \int^{\infty}_{0}
{d \nu \over \pi} \left | {{\bf p} \times {\hat {\bf q}} \over m}
\right |^2 {\rm Im} \ D_{11} ({\bf q}, \nu) \cr
&\times [ \ n_0 (\nu) G^> ({\bf p} + {\bf q}, \omega + \nu)
+ (n_0 (\nu) + 1) G^> ({\bf p} + {\bf q}, \omega - \nu) \ ] \ ,
}
\eqno{(35)}
$$
where $n_0 (\nu) = 1 / (e^{\nu / T} - 1)$ is the equilibrium boson
distribution function.
The real part of the retarded self-energy is given by
$$
\eqalign{
{\rm Re} \ \Sigma^R ({\bf p}, \omega; {\bf q}, \Omega) &=
- \int {d \omega' \over \pi} \ {\cal P}
{{\rm Im} \ \Sigma^R ({\bf p}, \omega'; {\bf q}, \Omega)
\over \omega - \omega'} \cr
&= - \int {d \omega' \over 2 \pi i} \
{\cal P} {\Sigma^> ({\bf p}, \omega'; {\bf q}, \Omega) -
\Sigma^< ({\bf p}, \omega'; {\bf q}, \Omega)
\over \omega - \omega'} \ ,
}
\eqno{(36)}
$$
where $\cal P$ represents the principal value and
${\rm Im} \ \Sigma^R  = {1 \over 2 i} (\Sigma^> - \Sigma^<)$ is
used.
The same relations hold for the Green's function $G^R$,
$$
{\rm Re} \ G^R ({\bf p}, \omega; {\bf q}, \Omega) =
- \int {d \omega' \over 2 \pi i} \
{\cal P} {G^> ({\bf p}, \omega'; {\bf q}, \Omega) -
G^< ({\bf p}, \omega'; {\bf q}, \Omega)
\over \omega - \omega'}
\eqno{(37)}
$$
and ${\rm Im} \ G^R = {1 \over 2 i} (G^> - G^<)$.

At equilibrium, the Green's functions $G^<$, $G^>$ can be written
as [26-28]
$$
\eqalign{
G^< ({\bf p}, \omega) &= i f_0 (\omega) A ({\bf p}, \omega) \ , \cr
G^> ({\bf p}, \omega) &= - i (1 - f_0 (\omega)) A ({\bf p}, \omega) \ ,
}
\eqno{(38)}
$$
where $A ({\bf p}, \omega)$ is given by Eq.(15).
{}From these relations, the one-loop self-energy $\Sigma^R_0$ at
equilibrium can be written as
$$
\Sigma^R_0 ({\bf p}, \omega) = \sum_{\bf q} \int^{\infty}_0
{d \nu \over \pi} \left | {{\bf p} \times {\hat {\bf q}} \over m}
\right |^2
\left [ \ {1 + n_0 (\nu) - f_0 (\xi_{{\bf p}+{\bf q}}) \over
\omega + i \delta - \xi_{{\bf p}+{\bf q}} - \nu} +
{n_0 (\nu) + f_0 (\xi_{{\bf p}+{\bf q}}) \over \omega + i \delta
- \xi_{{\bf p}+{\bf q}} + \nu} \ \right ]
\eqno{(39)}
$$
As emphasized in the previous section, if the self-energy depends
only on the frequency $\omega$, $A ({\bf p}, \omega)$ at equilibrium
is a peaked function of $\xi_p$.
Therefore, as far as the system is not far away from the equilibrium,
the generalized distribution function
$f (\theta_{\bf pq}, \omega; {\bf q}, \Omega)$, which is given
by the following relations, can be well defined at
zero temperature [25]:
$$
\eqalign{ \int {d \xi_p \over 2 \pi}
\left [ - i G^< ({\bf p}, \omega; {\bf q}, \Omega) \right ] &\equiv
f (\theta_{\bf pq}, \omega; {\bf q}, \Omega) \ , \cr
\int {d \xi_p \over 2 \pi}
\left [ i G^> ({\bf p}, \omega; {\bf q}, \Omega) \right ] &\equiv
1 - f (\theta_{\bf pq}, \omega; {\bf q}, \Omega) \ ,
}
\eqno{(40)}
$$
where $\theta_{\bf pq}$ is the angle between ${\bf p}$ and ${\bf q}$.

The extension to the case of finite temperatures requires special care
because, even at equilibrium, ${\rm Im} \ \Sigma^R_0 ({\bf p}, \omega)$
is known to be divergent [11]
so that $A ({\bf p}, \omega)$, $G^<_0$,
and $G^>_0$ at equilibrium are not well defined.
Therefore, the non-equilibrium $G^<$ and $G^>$ are also not well
defined near equilibrium.
In order to resolve this problem, let us first separate
the gauge field fluctuations into two parts, {\it i.e.},
${\bf a} ({\bf q}, \nu) \equiv {\bf a}_{-} ({\bf q}, \nu)$
for $\nu < T$ and
${\bf a} ({\bf q}, \nu) \equiv {\bf a}_{+} ({\bf q}, \nu)$
for $\nu > T$, then examine the effects of ${\bf a}_{+}$,
${\bf a}_{-}$ separately.
The classical fluctuation ${\bf a}_{-}$ of the gauge field can be
regarded as a vector potential which corresponds to a static but
spatially varying magnetic field
${\bf b}_{-} = \nabla \times {\bf a}_{-}$.
In order to remove the divergence in the self-energy,
one can consider the one-particle
Green's function ${\tilde G}_{-} \equiv
{\tilde G} ({\bf P}_{-}, \omega; {\bf r}, t)$
as a function of a new variable ${\bf P}_{-} = {\bf p} - {\bf a}_{-}$.
Since we effectively separate out ${\bf a}_{-}$ fluctuations,
the self-energy, which appears in the equation of motion given by Eq.(24),
should contain only ${\bf a}_{+}$ fluctuations and is free of divergences.
Therefore, $\delta G^<_{-} \equiv
\delta G^< ({\bf P}_{-}, \omega; {\bf r}, t)$ is well defined
and its equation of motion is given by the
Fourier transform of Eq.(30) with the following replacement.
In the first place, the variable ${\bf p}$ should be changed to
a new variable ${\bf P}_{-} = {\bf p} - {\bf a}_{-}$.
Secondly, the self-energy ${\tilde \Sigma}$ should be changed
to ${\tilde \Sigma}_{+}$ which contains now only ${\bf a}_{+}$
fluctuations.
Finally, the equation of motion contains a term
which depends on ${\bf b}_{-}$.
We argued in the appendix that ignoring this term does not affect
the physical interpretations of the QBE, which will appear in
sections IV, V, and VI.
We provide the details of the anaysis for the finite
temperature case in the appendix.
{}From now on, we will adopt the notation that
$G^<$ should be understood as
$G^<_{-}$ for finite temperatures.
For example, the generalized distribution function at finite
temperatures is given by Eq.(40) with the replacement that
$G^<, G^> \rightarrow G^<_{-}, G^>_{-}$.
The same type of abuse of notation applies to the self-energy,
where only ${\bf a}_{+}$ fluctuations should be included, {\it i.e.}
the QBE is valid at finite $T$, provided that
the lower cutoff $T$ is introduced for the frequency
integrals.

In Eq.(35), one can change the variables such that
${\bf p'} = {\bf p} + {\bf q}$ and $\omega' = \omega + \nu$.
The gauge field propagator can be written in terms of the new
variables as $D_{11} ({\bf q}, \nu) =
D_{11} ({\bf p'} - {\bf p}, \omega' - \omega)$, where $({\bf p}, \omega)$
and $({\bf p'}, \omega')$ represent the incoming and outgoing fermions.
Assuming that $|{\bf p}| \approx |{\bf p'}| \approx k_F$ and using
$|{\bf p'} - {\bf p}| \approx k_F |\theta_{\bf p'q} - \theta_{\bf pq}|$,
we get $D_{11} ({\bf q}, \nu) \approx
D_{11} (k_F |\theta_{\bf p'q} - \theta_{\bf pq}|, \omega' - \omega)$.
Using the above results and the fact that $G^<$ and $G^>$ are well
peaked functions of $\xi_p$ near the equilibrium,
${\rm Re} \ \Sigma^R$ can be written as
$$
{\rm Re} \ \Sigma^R  =
N (0) \int {d \theta_{\bf p'q} \over 2 \pi} \int d \omega' \
v^2_F \ {\rm Re} \ D_{11} (k_F |\theta_{\bf p'q} - \theta_{\bf pq}|,
\omega' - \omega) \ f (\theta_{\bf p'q}, \omega'; {\bf q}, \Omega) \ ,
\eqno{(41)}
$$
where $N (0) = {m \over 2 \pi}$ is the density of state.
Since we assume that the gauge field is at equilibrium,
$\delta ({\rm Re} \ \Sigma^R)$, which is the deviation
from the equilibrium, can be written as
$$
\delta ({\rm Re} \ \Sigma^R) =
N (0) \int {d \theta_{\bf p'q} \over 2 \pi} \int d \omega' \
v^2_F \ {\rm Re} \ D_{11} (k_F |\theta_{\bf p'q} - \theta_{\bf pq}|,
\omega' - \omega) \ \delta f (\theta_{\bf p'q}, \omega'; {\bf q}, \Omega) \ .
\eqno{(42)}
$$
We also assume that the non-equilibrium self-energy
depends only on $\omega$ as that of the equilibrium case,
which is plausible as far as the system is not far away
from the equilibrium.
In order to get the equation for
$f (\theta_{\bf pq}, \omega; {\bf q}, \Omega)$,
we perform $\int d \xi_p / 2 \pi$ integral on both sides of the
Eq.(30).
Note that
$$
\eqalign{&\int {d \xi_p \over 2 \pi} \ {\rm Re} \ G^R
({\bf p}, \omega'; {\bf q}, \Omega) \cr
&= \int {d \omega' \over 2 \pi} \ {\cal P}
{(1 - f (\theta_{\bf pq}, \omega'; {\bf q}, \Omega)) +
f (\theta_{\bf pq}, \omega'; {\bf q}, \Omega) \over \omega - \omega'} \cr
&= \int {d \omega' \over 2 \pi} \ {\cal P} {1 \over \omega - \omega'} = 0 \ .
}
\eqno{(43)}
$$
Thus the fourth and the fifth terms in the left hand side of the
QBE (given by Eq.(30)) vanish after $\int d \xi_p / 2 \pi$ integration.
After this integral, using Eqs.(36), (40) and (42), the remaining
parts of the Eq.(30) can be written as
($\delta f (\theta_{\bf pq}, \omega) \equiv
\delta f (\theta_{\bf pq}, \omega; {\bf q}, \Omega)$)
$$
\eqalign{
&[ \ \Omega - v_F q \ {\rm cos} \ \theta_{\bf pq} \ ] \
\delta f (\theta_{\bf pq}, \omega) \cr
&- N (0) \int {d \theta_{\bf p'q} \over 2 \pi} \int d \omega' \
v^2_F \ {\rm Re} \ D_{11} (k_F |\theta_{\bf p'q} - \theta_{\bf pq}|,
\omega' - \omega) \cr
&\hskip 3cm \times [ \ f_0 (\omega' + \Omega / 2) -
f_0 (\omega' - \Omega / 2) \ ] \
\delta f (\theta_{\bf pq}, \omega) \cr
&+ N (0) \int {d \theta_{\bf p'q} \over 2 \pi} \int d \omega' \
v^2_F \ {\rm Re} \ D_{11} (k_F |\theta_{\bf p'q} - \theta_{\bf pq}|,
\omega' - \omega) \cr
&\hskip 3cm \times [ \ f_0 (\omega + \Omega / 2) -
f_0 (\omega - \Omega / 2) \ ] \
\delta f (\theta_{\bf p'q}, \omega') \cr
&=
N(0) \int d \theta_{\bf p'q} \int^{\infty}_0 {d \nu \over \pi}
\int d \omega' \ v^2_F \ {\rm Im} \ D_{11}
(k_F |\theta_{\bf p'q} - \theta_{\bf pq}|, \nu) \cr
&\hskip 0.5cm \times \bigl [ \
\delta (\omega' - \omega + \nu) \
[ \ \delta f (\theta_{\bf pq}, \omega) \
(1 - f_0 (\omega') + n_0 (\nu) )
- \delta f (\theta_{\bf p'q}, \omega') \
(f_0 (\omega) + n_0 (\nu)) \ ] \cr
&\hskip 0.5cm - \delta (\omega' - \omega - \nu) \
[ \ \delta f (\theta_{\bf p'q}, \omega') \
(1 - f_0 (\omega) + n_0 (\nu))
- \delta f (\theta_{\bf pq}, \omega)
(f_0 (\omega') + n_0 (\nu)) \ ] \ \bigr ] \ .
}
\eqno{(44)}
$$

Some explainations of each term in the Eq.(44) are in order.
In the first place, as mentioned in the previous section,
the Eq.(44) is the analog of the usual QBE for
the quasi-particle distribution function
$\delta f ({\bf p}, {\bf q}, \Omega)$, thus the structures of
the QBEs in both cases are similar.
The first term on the left hand side of the equation corresponds
to the free fermions.
The second term on the left hand side corresponds to the self-energy
correction which renormalizes the mass of the fermions.
The third term on the left hand side can be regarded as the
contribution from the generalized Landau-interaction function
which can be defined as
$$
F (\theta_{\bf p'q} - \theta_{\bf pq}, \omega' - \omega)
= v^2_F \ {\rm Re} \
D_{11} (k_F |\theta_{\bf p'q} - \theta_{\bf pq}|, \omega' - \omega) \ .
\eqno{(45)}
$$
Note that this generalized Landau-interaction function contains
the frequency dependence as well as the usual angular dependence.
This is due to the fact that the gauge interaction is retarded
in time and it is also one of the major differences between the
fermion-gauge-field system and the usual Fermi liquid.
The right hand side of the equation is nothing but the collision
integral $I_{\rm collision}$ and is given by the Fermi-golden-rule.
Thus, Eq.(44) can be written as
$$
\eqalign{
&[ \ \Omega - v_F q \ {\rm cos} \ \theta_{\bf pq} \ ] \
\delta f (\theta_{\bf pq}, \omega) \cr
&- [ \ {\rm Re} \ \Sigma^R_0 (\omega + \Omega / 2)
- {\rm Re} \ \Sigma^R_0 (\omega - \Omega / 2) \ ] \
\delta f (\theta_{\bf pq}, \omega) \cr
&+ N (0) \int {d \theta_{\bf p'q} \over 2 \pi} \int d \omega' \
F (\theta_{\bf p'q} - \theta_{\bf pq}, \omega' - \omega) \
[ \ f_0 (\omega + \Omega / 2) -
f_0 (\omega - \Omega / 2) \ ] \
\delta f (\theta_{\bf p'q}, \omega') \cr
&= I_{\rm collision} \ .
}
$$

After taking the integral $\int d \omega / 2 \pi$ on both sides of Eq.(44),
it can be seen that one cannot write the QBE
only in terms of
$u (\theta_{\bf pq}, {\bf q}, \Omega) = \int d \omega / 2 \pi
\ \delta f (\theta_{\bf pq}, \omega; {\bf q}, \Omega)$
which is the generalized Fermi surface displacement.
That is, the QBE becomes
$$
\eqalign{
&[ \ \Omega - v_F q \ {\rm cos} \ \theta_{\bf pq} \ ] \
u (\theta_{\bf pq}, {\bf q}, \Omega) \cr
&- N (0) \int {d \theta_{\bf p'q} \over 2 \pi}
\int d \omega \int d \omega' \
v^2_F \ {\rm Re} \ D_{11} (k_F |\theta_{\bf p'q} - \theta_{\bf pq}|,
\omega' - \omega) \cr
&\hskip 2cm \times [ \ f_0 (\omega' + \Omega / 2) -
f_0 (\omega' - \Omega / 2) \ ] \
\bigl ( \delta f (\theta_{\bf pq}, \omega) -
\delta f (\theta_{\bf p'q}, \omega) \bigr ) \cr
&=
N(0) \int d \theta_{\bf p'q} \int^{\infty}_0 {d \nu \over \pi}
\int d \omega \int d \omega' \ v^2_F \ {\rm Im} \ D_{11}
(k_F |\theta_{\bf p'q} - \theta_{\bf pq}|, \nu) \cr
&\hskip 0.5cm \times [ \
\delta (\omega' - \omega + \nu) \
(1 - f_0 (\omega') + n_0 (\nu))
+ \delta (\omega' - \omega - \nu) \
(f_0 (\omega') + n_0 (\nu)) \ ] \cr
&\hskip 0.5cm \times \bigl (\delta f (\theta_{\bf pq}, \omega) -
\delta f (\theta_{\bf p'q}, \omega) \bigr ) \ .
}
\eqno{(46)}
$$
In the presence of the external potential $U ({\rm q}, \Omega)$,
one should add an additional term $v_F q \ {\rm cos} \ \theta_{\bf pq}
\ U ({\bf q}, \Omega)$ in the left hand side of Eq.(46), which requires
a careful derivation.
Note that the contributions from the self-energy and
the generalized Landau-interaction-function are combined
in the left hand side of the QBE.
Even though the above equation is already useful, it is worthwile
to transform this equation to the more familar one.
In the next section, we provide the approximate QBE for
$u (\theta_{\bf pq}, {\bf q}, \Omega)$ which is more useful
to understand the low energy excitations of the system.

\vskip 0.5cm

\centerline{\bf IV. QUANTUM BOLTZMANN EQUATION}
\centerline{\bf FOR THE GENERALIZED FERMI SURFACE DISPLACEMENT}

In order to transform the QBE given by Eq.(45) or Eq.(46) to a more
familiar form, it is necessary to simplify the generalized
Landau-interaction-function
$F (\theta, \omega) =
v^2_F \ {\rm Re} \ D_{11} (k_F |\theta|, \omega)$.
Note that
$$
{\rm Re} \ D_{11} (k_F |\theta|, \omega) =
{(\chi / \gamma^2) \ k^{2 + \eta}_F \ |\theta|^{2 + \eta} \over
\omega^2 + (\chi / \gamma)^2 \ k^{2 + 2 \eta}_F \
|\theta|^{2 + 2 \eta}} \ .
\eqno{(47)}
$$
It can be checked from Eq.(46) that
$\delta f (\theta_{\bf pq}, \omega; {\bf q}, \Omega)$ is
finite only when $|\omega| \le \Omega$ at zero temperature.
Therefore, the frequency $\omega$ in
${\rm Re} \ D_{11} (k_F |\theta|, \omega)$ is cutoff by $\Omega$.
In this case, one can introduce the $\Omega$ dependent cutoff
$\theta_c \approx {1 \over k_F}
\left ( {\gamma \Omega \over \chi} \right )^{1 \over 1 + \eta}$
in the angle variable and approximate $F (\theta, \omega)$ by the
following $F_{\rm Landau} (\theta)$.
$$
F_{\rm Landau} (\theta) =
\cases{F (\theta, \omega = 0) \ , &if $|\theta| > \theta_c$ \ ; \cr
F (\theta = \theta_c, \omega = 0) \ , &otherwise \ , \cr}
\eqno{(48)}
$$
where
$$
F (\theta, \omega = 0) = {v^2_F \over \chi k^{\eta}_F} \
{1 \over |\theta|^{\eta}} \ .
\eqno{(49)}
$$

Using this approximation and $f_0 (\omega) = \Theta (-\omega)$
at zero temperature, the QBE given by Eq.(46) at zero temperature
can be transformed into (the finite temperature case is discussed
in the appendix)
$$
\eqalign{
&[ \ \Omega - v_F q \ {\rm cos} \ \theta_{\bf pq} \ ] \
u (\theta_{\bf pq}, {\bf q}, \Omega) \cr
&+ \Omega \ N(0) \int {d \theta_{\bf p'q} \over 2 \pi} \
F_{\rm Landau} (\theta_{\bf p'q} - \theta_{\bf pq}) \
\bigl ( \ u (\theta_{\bf pq}, {\bf q}, \Omega) -
u (\theta_{\bf p'q}, {\bf q}, \Omega) \ \bigr ) \cr
&=
N(0) \int d \theta_{\bf p'q} \int^{\infty}_0 {d \nu \over \pi}
\int d \omega \int d \omega' \ v^2_F \ {\rm Im} \ D_{11}
(k_F |\theta_{\bf p'q} - \theta_{\bf pq}|, \nu) \cr
&\hskip 0.5cm \times [ \
\delta (\omega' - \omega + \nu) \
(1 - f_0 (\omega'))
+ \delta (\omega' - \omega - \nu) \
f_0 (\omega') \ ] \ \bigl (
\delta f (\theta_{\bf pq}, \omega) -
\delta f (\theta_{\bf p'q}, \omega) \bigr ) \ .
}
\eqno{(50)}
$$
Note that $\Omega \ N(0) \int d \theta_{\bf p'q} / 2 \pi \
F_{\rm Landau} (\theta_{\bf p'q} - \theta_{\bf pq}) \propto
\Omega^{2 \over 1 + \eta} \ (1 < \eta \le 2)$ or
$\Omega \ {\rm ln} \ \Omega \ (\eta = 1)$ corresponds to the
contribution from the real part of the retarded self-energy.
On the other hand,
$\Omega \ N(0) \int d \theta_{\bf p'q} / 2 \pi \
F_{\rm Landau} (\theta_{\bf p'q} - \theta_{\bf pq}) \
u (\theta_{\bf p'q}, {\bf q}, \Omega)$ represents the
Landau-interaction part.

For smooth fluctuations of the generalized Fermi surface
displacement, $u (\theta, {\bf q}, \Omega)$ is a
slowly varying function of $\theta$
so that there is a forward scattering cancellation between
the self-energy part and the Landau-interaction part.
Therefore, for smooth fluctuations,
the singular behavior of the self-energy
does not appear in the dynamics of the generalized
Fermi surface displacement.
One the other hand, for rough fluctuations,
$u (\theta, {\bf q}, \Omega)$ is a fastly varying
function.
In this case, the Landau-interaction part becomes very
small and the self-energy part dominates.
Thus, for rough fluctuations, the dynamics of
the generailzed Fermi surface displacement should show
the singular behavior of the self-energy.
{}From these results, one can expect that the smooth
and the rough fluctuations provide very different
physical pictures for the elementary excitations of the system.

One can make this observation more concrete by looking at
the QBE in angular momentum $l$ (which is the conjugate
variable of $\theta$) space.
By the following Fourier expansion,
$$
u (\theta, {\bf q}, \Omega) =
\sum_{l} \ e^{i l \theta} \ u_{l} ({\bf q}, \Omega) \ \ {\rm and} \ \
\delta f (\theta, \omega; {\bf q}, \Omega) =
\sum_{l} \ e^{i l \theta} \ \delta f_{l} (\omega; {\bf q}, \Omega) \ ,
\eqno{(51)}
$$
one can get
$$
\eqalign{
&\Omega \ u_{l} ({\bf q}, \Omega) -
{v_F q \over 2} \ [ \ u_{l+1} ({\bf q}, \Omega)
+ u_{l-1} ({\bf q}, \Omega) \ ] \cr
&+ \Omega \ N(0) \int {d \theta \over 2 \pi} \
F_{\rm Landau} (\theta) \
\bigl ( 1 - {\rm cos} \ (l \theta) \bigr ) \
u_{l} ({\bf q}, \Omega) \cr
&=
N(0) \int d \theta \int^{\infty}_0 {d \nu \over \pi}
\int d \omega \int d \omega' \ v^2_F \ {\rm Im} \ D_{11}
(k_F |\theta|, \nu) \
\bigl ( 1 - {\rm cos} \ (l \theta) \bigr ) \cr
&\hskip 0.5cm \times [ \
\delta (\omega' - \omega + \nu) \
(1 - f_0 (\omega'))
+ \delta (\omega' - \omega - \nu) \
f_0 (\omega') \ ] \
\delta f_{l} (\omega; {\bf q}, \Omega) \ .
}
\eqno{(52)}
$$

Note that, in the $1 - {\rm cos} \ (l \theta)$
factor inside the integral on the left hand side of
the QBE given by Eq.(52),
$1$ comes from the self-energy part and ${\rm cos} \ (l \theta)$
comes from the Landau-interaction part.
For $l < l_c \approx 1 / \theta_c \propto
\Omega^{-{1 \over 1 + \eta}}$,
$1 - {\rm cos} \ (l \theta) \approx l^2 \theta^2 / 2$ and
the additional $\theta^2$ dependence makes the angle integral
less sigular because typical $\theta$ is of the order of
$\Omega^{1 \over 1 + \eta}$.
Due to this cancellation for the small angle (forward)
scattering, the correction from the self-energy part
and the Landau-interaction part becomes of the order of
$\Omega^{4 \over 1 + \eta}$ so that it does not
cause any sigular correction.
Note that a similar type of cancellation occurs in the
collision integral.
Therefore, for the small angular momentum modes $l < l_c$,
the system behaves like the usual Fermi liquid.
For $l > l_c$, the ${\rm cos} \ (l \theta)$ factor becomes
highly oscillating as a function of $\theta$ so that
the Landau-interaction part becomes very small.
As a result, the self-energy part dominates and
the dispersion relation for the dynamics of the generalized
Fermi surface displacement is changed from
$\Omega = v_F q$ to $\Omega \propto q^{1 + \eta \over 2}$
($1 < \eta \le 2$) or $\Omega \propto q / |{\rm ln} \ q|$
($\eta =1$).
Also, a similar thing happens in the collision integral,
{\it i.e.}, the ${\rm cos} \ (l \theta)$ factor does not
contribute and the remaining contribution shows
the singular behavior of the imaginary part of the
self-energy so that the collision integral cannot be
ignored for $1 < \eta \le 2$ and can be {\it marginally}
ignored for $\eta = 1$.

Using the above results, one can understand the density-density
and the current-current correlation functions which show no
anomalous behavior in the long wave length and the
low frequency limits [12,13].
{}From the QBE, one can evaluate these correlation functions by
taking the angular average of the density or the current disturbance
due to the external potential and calculating the linear
response. As a result, in these correlation functions,
the small angular momentum modes are dominating so that
the results do not show any singular behavior.
{}From these results, one can also expect that two different
behaviors of the small ($l < l_c$) and the large ($l > l_c$)
angular momentum modes may show up even in the presence
of the finite effective magnetic field $\Delta B$ and
the large angular momentum modes may be responsible for
the singular energy gap of the system [6,14,15], which
is the subject of the next section.

\vskip 0.5cm

\centerline{\bf V. QUANTUM BOLTZMANN EQUATION IN THE PRESENCE OF}
\centerline{\bf EFFECTIVE MAGNETIC FIELD AND
THE ENERGY GAP}

We follow H\"ansch and Mahan [31] to derive the QBE in the presence of
the finite effective magnetic field $\Delta B$.
The only difference between the case of $\Delta B \not= 0$ and
that of $\Delta B = 0$ is that the external
vector potential $\Delta {\bf A} = - {1 \over 2} {\bf r} \times
\Delta {\bf B}$ enters to the knietic energy in the equation of
motion of the one-particle Green's function [31].
The same procedure used in the case of $\Delta B = 0$
can be imployed to derive the QBE from
the equation of motion of the one-particle Green's function.
The resulting equation can be transformed to a convenient
form by a change of variables given by
$$
{\bf P} = {\bf p} - \Delta {\bf A} = {\bf p} +
{1 \over 2} {\bf r} \times \Delta {\bf B}
\eqno{(53)}
$$
so that one can construct the QBE for
$G^< ({\bf P}, \omega; {\bf q}, \Omega)$ which is now
a function of ${\bf P}$ [31].
As a result, the change we have to make for the case of
$\Delta B \not = 0$ (compared to the case of $\Delta B = 0$
given by Eq.(30))
is that all the momentum variables should be changed from
${\bf p}$ to ${\bf P}$ and
the following additional terms should be added to Eq.(30) [31].
$$
\eqalign{
&{{\bf P} \over m} \cdot \Delta {\bf B} \times
{\partial \over \partial {\bf P}}
\delta G^< ({\bf P}, \omega; {\bf q}, \Omega)
+ {\partial \over \partial {\bf P}}
\delta \bigl ( {\rm Re} \
\Sigma^R ({\bf P}, \omega; {\bf q}, \Omega) \bigr )
\cdot \Delta {\bf B} \times {\partial \over \partial {\bf P}}
G^<_0 ({\bf P}, \omega) \cr
&- \Delta {\bf B} \cdot {\partial \over \partial {\bf P}}
\delta \Sigma^< ({\bf P}, \omega; {\bf q}, \Omega)
+ \Delta {\bf B} \cdot {\partial \over \partial {\bf P}}
\Sigma^<_0 ({\bf P}, \omega; {\bf q}, \Omega) \times
{\partial \over \partial {\bf P}} \delta \bigl (
{\rm Re} \ G^R ({\bf P}, \omega; {\bf q}, \Omega)
\bigr ) \ .
}
\eqno{(54)}
$$
Since the self-energy does not depend on the momentum {\bf P} in the
fermion-gauge-field system,
the only term which contributes to the QBE is
$$
{{\bf P} \over m} \cdot \Delta {\bf B} \times
{\partial \over \partial {\bf P}}
\delta G^< ({\bf P}, \omega; {\bf q}, \Omega) \ .
\eqno{(55)}
$$

In principle, the self-enegy and the Green's function in the QBE
also depend on the effective magnetic field $\Delta B$.
In the semiclassical approximation for very small $\Delta B$,
we ignore this type of $\Delta B$ dependence and, instead of that,
we introduce a low energy cutoff $E_{\rm g}$ in the frequency
integrals, which is the energy gap of the system.
Then, after the integration $\int d \xi_{\bf P} / 2 \pi$, the equation
becomes that of Eq.(30) with a low energy cutoff $E_{\rm g}$ and
it also contains an additional term given by
$$
{\Delta \omega_c} \ {\partial \over \partial \theta_{\bf Pq}} \
\delta f (\theta_{\bf Pq}, \omega; {\bf q}, \Omega) \ ,
\eqno{(56)}
$$
where $\Delta \omega_c = \Delta B / m$.
After $\int d \omega / 2 \pi$, the QBE for a generalized Fermi
surface displacement can be written as
$$
\eqalign{
&[ \ \Omega - v_F q \ {\rm cos} \ \theta_{\bf Pq} \ ] \
u (\theta_{\bf Pq}, {\bf q}, \Omega)
- i \Delta \omega_c \ {\partial \over \partial \theta_{\bf Pq}} \
u (\theta_{\bf Pq}, {\bf q}, \Omega)
\cr
&+ \Omega \ N(0) \int {d \theta_{\bf P'q} \over 2 \pi} \
F_{\rm Landau} (\theta_{\bf P'q} - \theta_{\bf Pq}) \
\bigl ( \ u (\theta_{\bf Pq}, {\bf q}, \Omega) -
u (\theta_{\bf P'q}, {\bf q}, \Omega) \ \bigr ) \cr
&=
N(0) \int d \theta_{\bf P'q} \int^{\infty}_0 {d \nu \over \pi}
\int d \omega \int d \omega' \ v^2_F \ {\rm Im} \ D_{11}
(k_F |\theta_{\bf P'q} - \theta_{\bf Pq}|, \nu) \cr
&\hskip 0.5cm \times [ \
\delta (\omega' - \omega + \nu) \
(1 - f_0 (\omega'))
+ \delta (\omega' - \omega - \nu) \
f_0 (\omega') \ ] \ \bigl (
\delta f (\theta_{\bf Pq}, \omega) -
\delta f (\theta_{\bf P'q}, \omega) \bigr ) \ ,
}
\eqno{(57)}
$$
where a low energy cutoff $E_{\rm g}$ is introduced in the
frequency integrals.
In particular, the angle cutoff $\theta_c$ in $F_{\rm Landau} (\theta)$
should be changed from $\theta_c \approx {1 \over k_F}
\left ( {\gamma \Omega \over \chi} \right )^{1 \over 1 + \eta}$
($\Delta B = 0$) to
$\theta_c \approx {1 \over k_F}
\left ( {\gamma E_{\rm g} \over \chi}
\right )^{1 \over 1 + \eta}$ ($\Delta B \not= 0$) in the low
frequency $\Omega$ limit.

Now similar interpretations can be made as the case of $\Delta B = 0$.
For the smooth fluctuations ($l < l_c \approx 1 / \theta_c$), there is
a cancellation between the self-energy and the Landau-interaction parts.
As a result, we have a term which is the order of
$\Omega \ E^{3 - \eta \over 1 + \eta}_{\rm g}$ which can be ignored
compared to $\Omega$ because $E_{\rm g}$ is very small near $\nu = 1/2$
or $\Delta B = 0$.
Also, a similar thing happens in the collision integral.
Therefore, the QBE for the smooth fluctuations can be written as
$$
[ \ \Omega - v_F q \ {\rm cos} \ \theta_{\bf Pq} \ ] \
u (\theta_{\bf Pq}, {\bf q}, \Omega)
- i \Delta \omega_c \ {\partial \over \partial \theta_{\bf Pq}} \
u (\theta_{\bf Pq}, {\bf q}, \Omega) \approx 0 \ .
\eqno{(58)}
$$
On the other hand, for the rough fluctuations ($l > l_c$),
the self-energy part dominates and we have a contribution
which is of the order $\Omega \ E^{-{\eta - 1 \over \eta + 1}}_{\rm g}$
($1 < \eta \le 2$) or $\Omega \ |{\rm ln} \ E_{\rm g}|$ ($\eta = 1$).
Ignoring $\Omega$ term compared to
$\Omega \ E^{-{\eta - 1 \over \eta + 1}}_{\rm g}$ ($1 < \eta \le 2$) or
$\Omega \ |{\rm ln} \ E_{\rm g}|$ ($\eta = 1$)
and multiflying the factor $E^{\eta - 1 \over \eta + 1}_{\rm g}$
($1 < \eta \le 2$) or $1 / |{\rm ln} \ E_{\rm g}|$ ($\eta = 1$) on
both sides of the equation, we get
$$
[ \ \Omega - v^*_F q \ {\rm cos} \ \theta_{\bf Pq} \ ] \
u (\theta_{\bf Pq}, {\bf q}, \Omega)
- i \Delta \omega^*_c \ {\partial \over \partial \theta_{\bf Pq}} \
u (\theta_{\bf Pq}, {\bf q}, \Omega) = {\rm collision \ integral} \ ,
\eqno{(59)}
$$
where $v^*_F = k_F / m^*$, $\Delta \omega^*_c = \Delta B / m^*$, and
$m^* / m  \propto E^{-{\eta - 1 \over \eta + 1}}_{\rm g}$
($1 < \eta \le 2$) or $|{\rm ln} \ E_{\rm g}|$ ($\eta = 1$).

Let us consider two different types of wave packets created along
the Fermi surface.
Note that the revolution of these wave packets is governed by
two different frequencies $\Delta \omega_c$
and $\Delta \omega^*_c$.
The frequency of revolution of the broad wave packet
(see Fig.2 (a)) is given by $\Delta \omega_c$ because
it mainly consists of small angular momentum modes.
On the other hand, if we ignore the collision integral in the QBE,
the frequency of revolution of the narrow
wave packet (see Fig.2 (b)) is given by
$\Delta \omega^*_c$ because
it mainly contains the large angular momentum modes.
The energy gap of the system can be obtained by quantizing
the motion of revolution and taking the smallest quantized
frequency as the energy gap of the system.
Therefore, the energy gap of the system is given by
$E_{\rm g} = \Delta \omega^*_c \propto
\Delta B \ E^{\eta - 1 \over \eta + 1}_{\rm g}$ ($1 < \eta \le 2$)
or $\Delta B / |{\rm ln} \ E_{\rm g}|$ ($\eta = 1$).
Solving this self-consistent equation for $E_{\rm g}$,
we get
$$
E_{\rm g} \propto
\cases{|\Delta B|^{1 + \eta \over 2} \ , &if $1 < \eta \le 2$ \ ; \cr
{|\Delta B| \over |{\rm ln} \ \Delta B|} \ , &if $\eta = 1$ \ . \cr
}
\eqno{(60)}
$$
This result is the same as the self-consistent treatment of
HLR [6] and also the perturbative evaluation of the activation
energy gap in the finite temperature compressibility [14].
We see that the divergent effective mass shows up in the energy gap
$E_{\rm g}$. More detailed discussions of the low lying excitations
described by the QBE can be found in the next section.

\vskip 0.5cm

\centerline{\bf VI. COLLECTIVE EXCITATIONS}

Let us first study the collective excitations of the system with
$\Delta B = 0$ by looking at the QBE given by Eq.(52).
We ignore the collision integral for the time being and
discuss its influence later.
In the absence of the collision integral, Eq.(52) can be
considered as the Schr\"odinger equation of an equivalent
tight binding model in the angular momentum space.
It is convenient to rewrite Eq.(52) as
$$
\eqalign{
\Omega \ v_l &= {v_F q\over 2} \
\left [ \ {v_{l+1}\over \sqrt{g(l)g(l+1)}} +
{v_{l-1}\over \sqrt{g(l)g(l-1)}} \ \right ] \ , \cr
v_l &= \sqrt{g(l)} \ u_l \ , \cr
}
\eqno{(61)}
$$
where
$$
g(l, \Omega) = 1 +
\ N(0) \int {d \theta \over 2 \pi} \
F_{\rm Landau} (\theta) \
\bigl ( 1 - {\rm cos} \ (l \theta) \bigr ) \ .
\eqno{(62)}
$$
Eq.(61) describes a particle hopping in a 1D lattice with a `spatial'
dependent hopping amplitude $t_l\approx {v_F q\over 2 g(l)}$.
Note that $g(l)$ is of the order one for $l< l_c$ and
becomes much larger,
$g(l) \propto \Omega^{-{\eta-1 \over \eta+1}}$, when $l>l_c$.
Due to this type of `spatial' dependent hopping amplitude,
the eigenspectrum of Eq.(61) consists of two parts.
That is, there is a continuous spectrum near the center of the
band and a discrete spectrum in the tail of the band.
The descrete spectrum
appears above and below the continuous spectrum (See Fig. 3).
The boundary between these two different spectra is determined from
$\Omega = 2 t_{l\to \infty} \propto v_Fq \ \Omega^{\eta-1 \over \eta+1}$,
which self-consistantly generates a
singular dispersion relation $\Omega (\theta) \propto
q^{1 + \eta \over 2}$ ($1 < \eta \le 2$) or
$\Omega (\theta) \propto q / |{\rm ln} \ q|$ ($\eta = 1$).
Also, the tail of the band ends at $\Omega (\theta) =2t_1 \sim v_F q$.

One can map this energy spectrum to
the diagram for the excitations in the usual
$\Omega-q$ plane, which is given by the Fig.4.
Note that the continuum states ($l > l_c$) can be mapped
to the particle-hole continuum which exist below
$\Omega \propto q^{1 + \eta \over 2}$ ($1 < \eta \le 2$)
or $\Omega \propto q / |{\rm ln} \ q|$ ($\eta = 1$).
On the other hand, the bound states (the discrete spectrum)
($l < l_c$) can be
mapped to the collective modes which exist between
$\Omega \propto q^{1 + \eta \over 2}$ ($1 < \eta \le 2$),
$\Omega \propto q / |{\rm ln} \ q|$ ($\eta = 1$) and
$\Omega \sim v_F q$.
However, the distinction between these two different
elementary excitations is obscured by the presence of
the collision integral which provides the life time for the
excitations.
In particular, since $g(l, \Omega)$ does not
provide a sharp boundary between $l > l_c$ and $l < l_c$,
one expects a crossover from the particle-hole excitations
to the collective modes even in the absence of the
collision integral.

Now let us consider the case of $\Delta B \not= 0$ ({\it i.e.}, away
from $\nu = 1/2$ state).
In this case, Eq.(61) becomes (see also Eq.(57))
$$
\Omega \ v_l = {l \Delta \omega_c \over g(l)} v_l+{v_F q\over 2}
\left [ {v_{l+1}\over \sqrt{g(l)g(l+1)}}
+{v_{l-1}\over \sqrt{g(l)g(l-1)}} \right ] \ .
\eqno{(63)}
$$
When $g(l)=1$, one can write the solution of Eq.(63) (or Eq.(57)) as
$$
u (\theta_{\bf Pq}, {\bf q}, t) \propto e^{in \theta_{\bf Pq}-i \Omega t} \
e^{- i {v_F q \over \Delta \omega_c} \ {\rm sin} \
\theta_{\bf Pq}}
$$
with $\Omega = n \Delta \omega_c$.
Thus, we recover the well known spectrum
of degenerate Landau levels for free fermions.

When $g(l) \neq$ const., it is difficult to calculate the spectrum of Eq.(63).
However, using $g(l)=g(-l)$, we can show that the spectrum of Eq.(63) is
symmetric about $\Omega=0$, and $\Omega=0$ is always an eigenvalue of Eq.(63).
Also, for non-zero $\Delta \omega_c$, the spectrum is always discrete.

Note that, for small $q\ll l_c \Delta \omega_c/v_F$,
$u (\theta_{\bf Pq}, {\bf q}, t)$ corresponds
to a smooth fluctuation of the Fermi surface. While, for large
$q \gg l_c \Delta \omega_c/v_F$, even the smooth parts of
$u (\theta_{\bf Pq}, {\bf q}, t)$, around $\theta_{\bf Pq}=\pm \pi/2$,
correspond to a rough fluctuation, hence the whole function
$u (\theta_{\bf Pq}, {\bf q}, t)$ corresponds to a rough fluctuation.
Thus, we expect that the small $q$ modes and the large $q$ modes have
very different dynamics.
The small $q$ modes should be controled by the finite effective mass
and the large $q$ modes, the divergent mass.

To understand the behavior of the modes in more detail, in the following,
we present a semiclassical calculation. The main result that we obtain
is the Eq. (75). The dispersion of the lowest lying mode
(for $q> \Delta \omega_c/v_F$) has a scaling form
$ \omega_{\rm cyc} (q)
\propto (\Delta \omega_c)^{1+\eta \over 2} f (q / q_c) $ with
$ f (\infty) = \ {\rm const.}$ and $f (x \ll 1) \propto x^{1-\eta}$.
The crossover momentum $q_c \propto \sqrt{\Delta \omega_c}$.

When $qv_F \ll \Delta \omega_c$ the spectrum can be calculated exactly and
is given by
$$
\Omega={l \Delta \omega_c \over g(l)} \ .
\eqno{(64)}
$$
To obtain the spectrum for $qv_F > \Delta \omega_c$
we will use a semiclassical approach.
Note that $(\theta_{\bf Pq}, l)$ is a canonical coordinate and momentum
pair. The classical Hamiltonian that corresponds to the quantum system
Eq.(63) can be found to be
$$
H(\theta_{\bf Pq}, l) =
{l \Delta \omega_c \over g(l)} +
{v_F q\over g(l)}{\rm cos}(\theta_{\bf Pq}) \ .
\eqno{(65)}
$$
Assuming $g(l)$ is a slowly varying function of $l$, one arrives
at the following simple classical equations of motion
$$
\dot \theta_{\bf Pq}={ \Delta \omega_c \over g(l)}, \ \ \ \ \
\dot l={v_F q\over g(l)}{\rm sin}(\theta_{\bf Pq}) \ .
\eqno{(66)}
$$
{}From this equation, one can easily show that
$$
l=-{v_F q\over \Delta \omega_c}{\rm cos}(\theta_{\bf Pq})+l_0 \ ,
\eqno{(67)}
$$
where $l_0$ is a constant.
Note that Eq.(67) with $l_0 = 0$ is an exact solution for the
classical system Eq.(65), which describes a motion with zero energy.
Now the first equation in Eq.(66) can be simplified as
$$
\dot \theta_{\bf Pq}={ \Delta \omega_c \over
g(-{v_F q\over \Delta \omega_c}{\rm cos}(\theta_{\bf Pq})+l_0)} \ ,
\eqno{(68)}
$$
which describes a periodic motion. The angular frequency of the periodic
motion is given by
$$
\omega={ 2\pi \Delta \omega_c  \over
\int_0^{2\pi} g(-{v_F q\over \Delta \omega_c}{\rm cos}(\theta_{\bf Pq})+l_0)
\ d\theta_{\bf Pq} } \ .
\eqno{(69)}
$$

The above classical frequency $\omega$ has a quantum interpretation.
It is the gap between neighboring energy levels, of which the
energy is close
to the classical energy associated with the classical motion described by
Eq.(67). In particular, the cylcotron frequency $\omega_{\rm cyc}$
is given by the gap between the $\Omega=0$ level and the first
$\Omega >0$ level.
Therefore
$$
\omega_{\rm cyc}={ 2\pi \Delta \omega_c  \over
\int_0^{2\pi} g(-{v_F q\over \Delta \omega_c}{\rm cos}(\theta_{\bf Pq})+1)
\ d\theta_{\bf Pq} } \ .
\eqno{(70)}$$
Here we have chosen $l_0 = 1$ (instead of $l_0 = 0$)
so that Eq.(70) repreduces the exact result
Eq.(64) when $q=0$. Note that $g(l)$ also depends on frequency $\Omega$
and we should set $\Omega=\omega_{\rm cyc}$ in the function $g(l)$.
Thus, the cylcotron frequency should be self-consistantly determined from
Eq.(70).

We would like to remark that when $q\gg \Delta \omega_c/v_F$, the classical
frequency in Eq. (69) is a smooth function of $l_0$, hence
a smooth function of the energy. This means that
the gap between the neighboring energy levels is also a smooth function
of the energy of the levels. The validity of the semiclassical approach
requires that the gap between neighboring energy levels is almost
a constant in the neighborhood of interested energies. Thus
the above behavior of the classical frequency indicates that
the semiclassical approach is at least self-consistant.

To analyze the behavior of $\omega_{\rm cyc}$, we first make an
approximation for Eq.(70) as
$$
\omega_{\rm cyc}={  \Delta \omega_c  \over
g(\lambda {v_F q\over \Delta \omega_c}+1) } \ ,
\eqno{(71)}
$$
where $\lambda$ is a non-zero constant between 0 and 1.
We see that $\omega_{\rm cyc}(q)$ has a sharp dependence on $q$ around
$q \sim \Delta \omega_c / v_F$.
The smaller the  $\Delta \omega_c$ the sharper
the $q$ dependence.
This sharp dependence is not due to the singular
gauge interaction, but merely a consequence
of the fact that $g(1) \neq g(2) \neq \cdots$.

As $q$ increases, $g(\lambda {v_F q \over \Delta \omega_c}+1)$
becomes larger and larger,
thus we expect that $\omega_{\rm cyc}(q)$ decreases.
When $q$ exceeds a crossover value
$q_c$, $g(\lambda {v_F q\over \Delta \omega_c}+1)$
saturates at a very large value and $\omega_{\rm cyc} (q)$ is
drastically reduced.
This phenomena is a result of the singular gauge interaction.
The crossover momentum $q_c$ is determined from
$$
{v_F q_c\over \Delta \omega_c}=l_c
= k_F \left ( {\chi \over \gamma \omega_{\rm cyc} (q \rightarrow \infty)}
\right )^{1 \over 1+\eta} \ ,
$$
and
$$
\eqalign{
\omega_{\rm cyc} (q \rightarrow \infty) &= {\Delta \omega_c
\over C (\eta)
(\omega_{\rm cyc} (q \rightarrow \infty))^{1-\eta \over 1+\eta}}
\ {\rm for} \ 1 < \eta \le 2, \cr
\omega_{\rm cyc} (q \rightarrow \infty) &= {\Delta \omega_c
\over C (\eta = 1) |{\rm ln} \ \omega_{\rm cyc} (q \rightarrow \infty)|}
\ {\rm for} \ \eta = 1 \ ,
}
\eqno{(72)}
$$
where
$$
C (\eta) = {v_F \ {\rm cos} \
\left [ {\pi \over 2} \left ( {\eta - 1 \over \eta + 1} \right ) \right ]
\over
2 \pi (1 + \eta) \ {\rm sin} \ \left ( {2 \pi \over 1 + \eta} \right )
\gamma^{\eta - 1 \over \eta + 1} \chi^{2 \over 1 + \eta}}
$$
for $1 < \eta \le 2$ and $C (\eta = 1) = {v_F \over 2 \pi^2 \chi}$ for
$\eta = 1$.
We find
$$
\eqalign{
q_c &= B (\eta) \sqrt{\Delta \omega_c} \ \ \ \ \hbox{for }
1 < \eta \leq 2 \ , \cr
q_c &= B (\eta = 1) \sqrt{\Delta \omega_c |{\rm ln} \ \Delta \omega_c|}
\ \ \ \ \hbox{for } \eta = 1 \ ,}
\eqno{(73)}
$$
where $B (\eta) = m (\chi / \gamma)^{1 \over 1 + \eta} \sqrt{C(\eta)}$.
When $q \gg q_c$, the cyclotron frequency saturates at the following
values.
$$\eqalign{
\omega_{\rm cyc} (q \rightarrow \infty)
&= (\Delta \omega_c / C (\eta))^{1+\eta\over 2}
\ \ \ \ \hbox{for } 1 <\eta \leq 2 \ , \cr
\omega_{\rm cyc} (q \rightarrow \infty)
&= {\Delta \omega_c / C (\eta = 1)
\over |{\rm ln} (\Delta \omega_c / C (\eta = 1))|}
\ \ \ \ \hbox{for } \eta = 1 \ .}
\eqno{(74)}$$
When $v_F q / \Delta \omega_c \gg 1$, the cyclotron freqency is expected
to have the following scaling form:
$$
\eqalign{
\omega_{\rm cyc} (q)
&\propto (\Delta \omega_c)^{1+\eta \over 2}
f (q / q_c) \ , \cr
f (\infty) &= \ {\rm const.} \ \ \ \ {\rm and } \ \ \ \
f (x \ll 1) \propto x^{1-\eta} \ ,}
\eqno{(75)}
$$
where $f (\infty)$ is determined from
$\omega_{\rm cyc} (q \rightarrow \infty) \propto
(\Delta \omega_c)^{1+\eta\over 2}$ and
$f (x \ll 1)$ can be obtained from the condition that
$\omega_{\rm cyc} (q) = \Delta \omega_c$ for
$q \sim \Delta \omega_c / v_F$.
Note that the divergence of $f(x)$ for small  $x$ should be
cutoff when $x \sim \Delta \omega_c / v_F q_c$.
As a result, the cyclotron spectrum of the system looks
like the one given by Fig.5.

The smaller gap for $q > q_c$ corresponds to a divergent effective mass
$m^*\propto ( \Delta \omega_c)^{1-\eta\over 1+\eta}$, while the larger
gap near $q = 0$ can be viewed as a cyclotron freqency derived from a finite
effective mass.
The thermal activation gap measured through the longitudinal
conductance is given by the smaller gap at large wave vectors $q > q_c$.
However the cyclotron frequency measured through the cyclotron resonance
for the uniform electric field should be given by the larger gap.

The above discussion of the cyclotron frequency is for the toy model, where
only the transverse gauge field fluctuations are included.
One may wonder whether the same picture also applies to the real
$\nu = 1/2$ state.
In the real $\nu = 1/2$ state, the lowest lying plasma modes correspond
to the intra-Landau-level excitations, of which energy is much less than
the inter-Landau-level gap $\omega_c$.
In the $q \to 0$ limit, such modes decouple
from the center of mass motion.
This means that the $u_{\pm 1}$ components
(which correspond to the dipolar distortions of the Fermi surface)
of the eigenmodes must disappear in the $q \to 0$ limit as far as the
lowest lying modes (intra-Landau-level modes) are concerned.
The mode that contains
$u_{\pm 1}$ components should have the big inter-Landau-level gap
in the $q \to 0$ limit in order to satisfy the Kohn's theorem.
Examining our solution for the eigenmodes in the $q \to 0$ limit,
we find that the lowest eigenmodes are given by
$u_l \propto \delta_{\pm 1,l}$.
Therefore, according to the above consideration,
we cannot identify the lowest lying modes in the toy model with
the lowest lying intra-Landau level plasma modes in the real model.
However, this problem can be fixed following the
procedure introduced in Ref.30.
That is, we may introduce an additional non-divergent
Landau-Fermi-liquid parameter $\Delta F_1$ which
modifies only the value of $g(\pm 1)$.
We may fine-tune the value of $\Delta F_1$
such that the $l = \pm 1$ modes in Eq.(64)
will have the big inter-Landau-level gap
$\Omega={\Delta \omega_c \over g(\pm 1)}=\omega_c$.
In this case the
$l=\pm 2$ modes become the lowest lying modes in the $q \to 0$ limit.
Such modes correspond to the quadradpolar distortions of the Fermi
surface and decouple from the center of mass motion.
The above correction only affects the energy of the lowest lying modes
for the small momenta, $q < \Delta \omega_c/v_F$.
With this type of correction, our results for the toy model
essentially applies to the $\nu = 1/2$ state.
The only change is that the lowest lying modes at small momenta,
$q \ll \Delta \omega_c/v_F$, is given by the $l=\pm 2$ modes instead of
the $l=\pm 1$ modes. This is because as $q$ decreases below a value
of order $\Delta \omega_c/v_F$, the $l=\pm 1$ modes start to have
a higher energy than that of the $l=\pm 2$ modes,
and the lowest lying modes crossover to the $l=\pm 2$ modes.

In the absence of the singular gauge interaction, according to the
picture developed in Ref.30, one expects that
the intra-Landau-level plasma mode
near $\nu=1/2$ has a gap $2 \Delta \omega_c$ for
$q < \Delta \omega_c / v_F$.
The gap is expected to be reduced by the factor 2 when
$q > \Delta \omega_c / v_F$.
In the presence of the singular gauge interaction, we find that
the plasma mode has a gap of order
$2 \Delta \omega_c$ (since $g(\pm 2) \neq 1$) for
$q < \Delta \omega_c/v_F$.
However, the gap for the large momenta can be much less than
$\Delta \omega_c$.
Observing this drastic gap reduction will confirm the presence of
the singular gauge interaction.

In the above disscusion, we have ignored the effects of the collision term.
The role of collision integral is simply to provide the life time effects
on the collective excitations. However, due to the energy conservation, only
the collective modes with energy greater than
$2 \omega_{\rm cyc} (q_{\rm min})$ will have a finite life time.
Here $\omega_{\rm cyc}(q_{\rm min})$ is the minimum
energy gap of the lowest lying plasma mode and $q_{\rm min}$ is
the momentum where the energy takes the minimum value.
For large $q$, the modes above $2 \omega_{\rm cyc} (q_{\rm min})$ may
have a short life time such that the modes are not well defined.

\vskip 0.5cm

\centerline{\bf VII. SUMMARY, CONCLUSION,}
\centerline{\bf AND IMPLICATIONS TO EXPERIMENTS}

In this section, we summarize the results and provide the
unified picture for the composite fermions interacting with
a gauge field.
In this paper, we construct a general framework, which is the
QBE of the system, to understand
the previously known theoretical [6,12-16] and
experimental [1-3,8-10] results.
Since there is no well defined Landau-quasi-particle,
we cannot use the usual formulation of the QBE so that
we used an alternative formulation which was used by
Prange and Kadanoff [25] for the electron-phonon problem.
We used the non-equilibrium Green's function technique [25-28] to
derive the QBE of the generalized distribution function
$\delta f (\theta_{\bf pq}, \omega; {\bf q}, \Omega)$ for
$\Delta B = 0$, and
$\delta f (\theta_{\bf Pq}, \omega; {\bf q}, \Omega)$
(${\bf P} = {\bf p} - \Delta {\bf A}$) for $\Delta B \not= 0$.
{}From this equation, we also derived the QBE for the generalized
Fermi surface displacement $u (\theta_{\bf pq}, {\bf q}, \Omega)$
($\Delta B = 0$) or $u (\theta_{\bf Pq}, {\bf q}, \Omega)$
($\Delta B \not= 0$)
which corresponds to the local
variation of the chemical potential in momentum space.

For $\Delta B = 0$, the QBE consists of three parts;
the self-energy part, the generalized Landau-interaction part,
and the collision integral.
The Landau-interaction function $F_{\rm Landau} (\theta)$ can
be taken as $F_{\rm Landau} (\theta) \propto 1 / |\theta|^{\eta}$
for $\theta > \theta_c \propto \Omega^{1 \over 1 + \eta}$ and
$1 / |\theta_c|^\eta$ for $\theta < \theta_c$.
For the smooth fluctuations of the genaralized Fermi surface
displacement
($l < l_c \approx 1 / \theta_c \propto
\Omega^{-{1 \over 1 + \eta}}$), where $l$
(the angular momentum in momentum space) is the
conjugate variable of the angle $\theta$, there is a
small-angle-(forward)-scattering cancellation between the self-energy
part and the Landau-interaction part.
Both of the self-energy part and the Landau-interaction part
are of the order $\Omega^{2 \over 1 + \eta}$ ($1 < \eta \le 2$)
or $\Omega \ {\rm ln} \ \Omega$ ($\eta = 1$).
After cancellation, the combination of these contributions becomes
of the order $\Omega^{4 \over 1 + \eta}$.
There is also a similar
cancellation in the collision interal so that the transport
scattering rate becomes of the order $\Omega^{4 \over 1 + \eta}$.
As a result, the smooth fluctuations show no anomalous behavior
expected from the singular self-energy correction.
On the other hand, for the rough fluctuations ($l > l_c$),
the Landau-interaction part becomes very small and
the self-energy part, which is proportional to
$\Omega^{2 \over 1 + \eta}$, dominates.
Also the collision integral becomes of the order
$\Omega^{2 \over 1 + \eta}$.
Therefore, the rough fluctuations show anomalous behavior
of the self-energy correction and suggest that the effective
mass shows a divergent behavior
$m^* \propto \Omega^{-{\eta - 1 \over \eta + 1}}$
for $1 < \eta \le 2$ and $m^* \propto |{\rm ln} \ \Omega|$ for
$\eta = 1$.

{}From these results, one can understand the density-density and
the current-current correlation functions calculated in the
perturbation theory [12,13], which show no anomalous behavior in the
long wave length and the low frequency limits.
Using the QBE, one can evaluate these correlation functions by
taking the angular average of the density or current disturbance
due to the external potential and calculating the linear
response. Thus, in these correlation functions, the small angular
momentum modes are dominating so that the results do not show any
singular behavior.
Note that the cancellation which exists in the collision integral
implies that the transport life time is sufficiently long to
explain the long mean free path of the composite fermions in the
recent magnetic focusing experiment [10].
For the $2 k_F$ response functions, there is no corresponding
cancellation between the self-energy part and
the Landau-interaction part
so that it shows the singular behavior [13].

The QBE in the presence of the small effective magnetic field
$\Delta B$ was used to understand the energy gap $E_{\rm g}$
of the system.
As the case of $\Delta B = 0$, there can be two different
behaviors of the generalized Fermi surface displacement.
For the smooth fluctuations
($l < l_c \propto E^{-{1 \over 1 + \eta}}_{\rm g}$),
the frequency of revolution of the wave packet is
given by $\Delta \omega_c = \Delta B / m$, ${\it i. e.}$,
there is no anomalous behavior after the cancellation
between the self-energy and the Landau-interaction parts.
For the rough fluctuations, the self-energy part dominates
and the frequency of revolution of the wave packet is
renormalized as
$\Delta \omega^*_c \propto \Delta \omega_c \
E_{\rm g}^{-{\eta - 1 \over \eta + 1}}$.
The energy gap can be obtained by quantizing the motion
of the wave packet and taking the lowest quantized
frequency which is nothing but $\Delta \omega^*_c$.
Solving the self-consistent equation
$E_{\rm g} = \Delta \omega^*_c$,
we get
$E_{\rm g} \propto |\Delta B|^{1 + \eta \over 2}$ for
$1 < \eta \le 2$ and
$E_{\rm g} \propto
{|\Delta B| \over |{\rm ln} \ \Delta B|}$ for $\eta = 1$.
These are consistent with the previous results [6,14,15].

The excitations of the system were studied from the
QBE of the generalized Fermi surface displacement.
For $\Delta B = 0$, in the absence of the collision integral,
there are two types of the excitations which can be
described most easily in the $\Omega - q$ plane.
There are particle-hole excitations which exist below
an edge
$\Omega \propto q^{1 + \eta \over 2}$ ($1 < \eta \le 2$)
or $\Omega \propto q / |{\rm ln} \ q|$ ($\eta = 1$).
There are also collective modes which exist between
$\Omega \propto q^{1 + \eta \over 2}$ ($1 < \eta \le 2$),
$\Omega \propto q / |{\rm ln} \ q|$ ($\eta = 1$) and
$\Omega \sim v_F q$.
However, the distinction between these two different
elementary excitations is obscured by the presence of
the collision integral which provides the life time of the
excitations.
In the case of $\Delta B \not= 0$, the QBE in the presence
of the finite $\Delta B$ is again used to understand
the low lying plasma spectrum of the system
as a function of $q$.
For $q < q_c$, where
$q_c \propto \sqrt{|\Delta B|}$ for $1 < \eta \le 2$ and
$q_c \propto \sqrt{|\Delta B| \ {\rm ln} \ |\Delta B|}$ for
$\eta = 1$, the plasma mode corresponds to a smooth fluctuation
of the Fermi surface, and the excitation gap is given by
$\Delta \omega_c \sim \Delta B / m$.
On the other hand, for $q > q_c$, the plasma mode corresponds to
a rough fluctuation of the Fermi surface. As a consequence,
 the excitation gap becomes
much smaller and proportional to $|\Delta B|^{1 + \eta \over 2}$ for
$1 < \eta \le 2$ and $|\Delta B| / |{\rm ln} \ \Delta B|$ for
$\eta = 1$.
Thus, the lowest excitation spectrum of the system looks
like the one given by Fig.5, which is consistent with the
previous numerical calculations [30].

Applying the picture developed in this paper for the $\nu = 1/2$
metallic state to the magnetic focusing experiment of Ref.10,
we find that the observed oscillations should not be interpreted as
the effects due to the focusing of the quarsiparticles.
This is because the inelastic mean free path
$L_q = v_F^* \tau$ and the life time $\tau\sim 1/T$ of the quasiparticle
is quite short.
Here $v_F^*$ is the renormalized Fermi velocity of the quasiparticle.
For the Coulomb interaction, we find
$$L_q\sim {\sqrt{ 4\pi n}\over m T{\rm ln} (E_F/T)}$$
Here $n$ is the density of the electron, $T$ the temperature, $m$
the bare mass of the composite fermion,
and $E_F={k_F^2\over 2 m}={2\pi n\over m}$.
Taking $n=10^{11} cm^{-2}$ and $m$ to be the electron mass in the vacuum
(see Ref.2, Ref.3 and Ref.10), we have
$$L_q \sim 0.26 {100{\rm mK} \over T } \mu{\rm m}$$
At $T=35$mK, $L_q \sim 0.7\mu$m which is much less than the length
of the semi-circular path, $6\mu$m, which connects the two slits.
Therefore, the oscillations observed in Ref.10 cannot be explained
by the focusing of the quasiparticles which have a divergent
effective mass and a short life time.

There is another way to explain the observed oscillations in Ref.10.
We can inject a net current into one slit, which causes a dipolar
distortion of the local Fermi surface near the slit.
The current and the associated
dipolar distortion propagate in space according to the QBE
and are bended by the effective
magnetic field $\Delta B$. This causes the oscillation in the current
received by the other slit. According to this picture, the oscillations
observed in Ref. 10 is caused by the smooth fluctuations of the Fermi
surface whose dynamics is identical to those of a Fermi liquid with
a {\it finite} effective mass.
Thus, the oscillations in the magnetic focusing experiments behave
as if they are caused by quasiparticles with a finite effective mass
and a long life time. The relexation time for the current distribution
is given by $\tau_j \sim {E_F \over T^2 \ {\rm ln} \ (E_F/T)}$ for the
Coulomb interaction.
This leads to a diffussion length (caused by the gauge fluctuations)
$L_j=v_F \tau_j$, where $v_F$ is the bare Fermi velocity of the composite
fermions. We find
$$L_j\sim 14 \left ( {100 {\rm mK} \over T} \right )^2 \mu{\rm m}
$$
The real diffussion length should be shorter than the above value due to
other possible scattering mechanisms.
Thus, we expect that the crossover temperature, above which the
oscillations disappear, should be lower than 150mK.
In the experiment [10], no oscillations
were observed above 100mK.
Another important consequence of our picture is that, if a time-of-flight
measurement can be performed by pulsing the incoming current,
the time is given by the bare velocity $v_F$ and {\it not} the
quasiparticle velocity $v^*_F$.

Finally, we make a remark on the surface acoustic wave experiment.
The condition that we can see the resonance between the
cyclotron radius and sound wave length
is given by
$\omega_{\rm cyc} \gg \omega_{\rm s}$, where $\omega_{\rm cyc}$
is the cyclotron frequency and $\omega_{\rm s}$ is the sound
wave frequency.
The reason is that we can regard the sound wave as a standing
wave only when $\omega_{\rm cyc} \gg \omega_{\rm s}$.
Let us imagine that we are changing $\omega_{\rm s}$ such that
$\omega_{\rm s} \approx \Delta \omega^*_c$.
If we use the quasiparticle picture to explain the above resonance,
then the cyclotron frequency
$\omega_{\rm cyc}$ is determined by the divergent effective
mass, and $\omega_{\rm cyc}$ should be comparable to
$\Delta \omega^*_c$.
Therefore, there should not be any resonance because
$\omega_{\rm cyc} \approx \omega_{\rm s}$ in this case.
However, in reality, the resonance is governed by the
smooth fluctuation of the Fermi surface, so that $\omega_{\rm cyc} \approx
\Delta \omega_c$ is a cyclotron frequency
determined by the finite bare mass of the composite fermion.
As a result, one should still see the resonance because
$\omega_{\rm cyc} \gg \omega_{\rm s} \approx \Delta \omega^*_c$.
Therefore, one can expect that there should be still resonance
effects even when the phonon energy exceeds the energy gap
determined from the
Shubnikov-de Haas oscillations.
The bottom line is that the cyclotron frequency measured in acoustic
wave experiments can be much larger than the energy gap measured in
transport experiments.
In a recent experiment of Willet {\it et. al} [32], resonance was
observed when $\omega_s$ is larger than the energy gap of the
system determined by the large effective mass obtained from the
Shubnikov-de Haas oscillations [3]. The authors claimed that
this is an apparent contradiction between the surface acoustic
wave experiment and the Shubnikov-de Haas oscillations.
We would like to point out that the cyclotron frequency
(for small $q$)
is determined by the bare mass
(In a crude estimation [6], the bare mass is about 1/3 of the
electron mass in vacuum).
On the other hand, the mass obtained from Shubunikov-de Haas
oscillations or from the activation gap in transport measurements
is in principle a different mass, which in practice turns out to be
of order of the electron mass in vacuum even away from $\nu = 1/2$.
Even though we do not understand quantitatively the mass difference,
there is in principle no contradiction.
The surface acoustic experiment is in fact an excellent way of
measuring the bare mass.

\vskip 0.5cm

\centerline{\bf ACKNOWLEDGMENTS}

We are grateful to A. Stern and B. I. Halperin for discussing their
results with us prior to the publication.
We also would like to thank A. Furusaki, W. Kang, A. J. Millis,
T. M. Rice, M. Sigrist, H. L. Stormer, R. L. Willet, A. Yacoby for
helpful discussions and
E. H. Fradkin, C. M. Varma, P. B. Wiegmann for enlightening comments.
YBK and XGW are supported by NSF
grant No. DMR-9411574.
PAL is supported by NSF grant No. DMR-9216007.

\vfill\vfill\vfill
\break

\vskip 0.5cm

\centerline{\bigbf Appendix}

\vskip 0.5cm

In this appendix, we consider the QBE at finite temperatures.
Recall that ${\rm Im} \ \Sigma^R ({\bf p}, \omega)$ at equilibrium
diverges at finite temperatures, which has no cutoff [11].
In this case, it is clear from Eq.(15) that
$G^<_0 ({\bf p}, \omega) = i f_0 (\omega) A ({\bf p}, \omega)$ is
not well defined. Thus, it is also difficult to define
$G^< ({\bf p}, \omega; {\bf r}, t)$ for the non-equilibrium case.
Since the divergent contribution to the self-energy comes from
the gauge field fluctuations with $\nu < T$, where $\nu$ is the
energy transfer by the gauge field [11], it is worthwhile to separate
the gauge field fluctuations into two parts, {\it i.e.},
${\bf a} ({\bf q}, \nu) \equiv {\bf a}_{-} ({\bf q}, \nu)$
for $\nu < T$ and
${\bf a} ({\bf q}, \nu) \equiv {\bf a}_{+} ({\bf q}, \nu)$
for $\nu > T$, and examine the effects of ${\bf a}_{+}$,
${\bf a}_{-}$ separately.

The classical fluctuation ${\bf a}_{-}$ of the gauge field can be
regarded as a vector potential which corresponds to a static but
spatially varying magnetic field
${\bf b}_{-} = \nabla \times {\bf a}_{-}$.
For a given random `magnetic' field, ${\bf b}_{-}({\bf r})$, and in
a fixed gauge, the fluctuation of the gauge potential ${\bf a}_{-}$
can be very large.
The gauge potential can have huge differences
from one point to another, as long as the two points are well separated.
We know that locally the center of the Fermi surface is at the momentum
${\bf p} - {\bf a}_{-} ({\bf r})$ around the point ${\bf r}$ in space.
The huge fluctuation of ${\bf a}_{-}$
indicates that the local Fermi surfaces at different points in space
may appear in very different regions in the momentum space.
This is the reason why the one-particle
Green's function in the {\it momentum space} is not well defined.
This also suggests that the Fermion distribution in the momentum space,
$f ({\bf p}, \omega)$, may be ill-defined.
Note that the local Fermi surface can be determined in terms of
the velocity of the fermions ({\it i.e.}, the states with
${m \over 2} {\bf v}^2 ={1\over 2m}({\bf p }-{\bf a}_-)^2 < E_F$
are filled) and the velocity is a gauge-invarint physical quantity.
This suggests that it is more reasonable to study the fermion distribution
in the physical {\it velocity space}.
The above discussion leads us to consider the one-particle
Green's function ${\tilde G} ({\bf P}_{-}, \omega; {\bf r}, t)$
as a function of a new variable ${\bf P}_{-} =m{\bf v}= {\bf p} - {\bf a}_{-}$.
Note that this transformation is reminicent of the procedure we used
in the case of the finite effective magnetic field (see section V).
We may follow the similar line of derivation to obtain the QBE in the
random magnetic field.
Since we effectively separate out ${\bf a}_{-}$ fluctuations,
the self-energy, which appears in the equation of motion given by Eq.(24),
should contain only ${\bf a}_{+}$ fluctuations.
Therefore, the equation of motion for
$\delta G^< ({\bf P}_{-}, \omega; {\bf r}, t)$ is given by the
Fourier transform of Eq.(30) with the following replacement.
In the first place, the variable ${\bf p}$ should be changed to
a new variable ${\bf P}_{-} = {\bf p} - {\bf a}_{-}$.
Secondly, the self-energy ${\tilde \Sigma}$ should be changed
to ${\tilde \Sigma}_{+}$ which contains now only ${\bf a}_{+}$
fluctuations.
Finally, as we can see from the case of the finite
effective magnetic field in section V, the following term
should be added:
$$
{{\bf P}_{-} \over m} \cdot {\bf b}_{-} ({\bf r}) \times
{\partial \over \partial {\bf P}_{-}}
\delta G^< ({\bf P}_{-}, \omega; {\bf r}, t) \ .
\eqno{({\rm A}.1)}
$$
Note that the equation of motion
contains the term which depends on ${\bf b}_{-}$, but does not contain
the terms which depend on ${\bf a}_{-}$ in an explicit way.
Since we removed the source of the divergence (non-gauge-invariance
with respect to ${\bf a}_{-}$),
the Green's function
${\tilde G} ({\bf P}_{-}, \omega; {\bf r}, t)$ or the corresponding
self-energy is now finite for finite $T$ or $\omega$.

Now one can perform the integration $\int d \xi_{{\bf P}_{-}} / 2 \pi$ of
$\delta G^< ({\bf P}_{-}, \omega; {\bf r}, t)$ safely to define
$$
\eqalign{ \int {d \xi_{{\bf P}_{-}} \over 2 \pi}
\left [ \ - i G^< ({\bf P}_{-}, \omega; {\bf r}, t) \ \right ] &\equiv
f (\theta, \omega; {\bf r}, t) \ , \cr
\int {d \xi_{{\bf P}_{-}} \over 2 \pi}
\left [ \ i G^> ({\bf P}_{-}, \omega; {\bf r}, t) \ \right ] &\equiv
1 - f (\theta, \omega; {\bf r}, t) \ , }
\eqno{({\rm A}.2)}
$$
where $\theta$ is the angle between ${\bf P}_{-}$ and a given direction.
For a while, let us ignore the contribution coming from the
term that depends on ${\bf b}_{-} ({\bf r})$ in the equation of
motion for
$\delta f (\theta, \omega; {\bf r}, t)$, which
is given by
$$
{b_{-} ({\bf r}) \over m} \ {\partial \over \partial \theta} \
\delta f (\theta, \omega; {\bf r}, t) \ .
\eqno{({\rm A}.3)}
$$
In the absence of this term,
the equation of motion of the generalized distribution function
$\delta f (\theta, \omega; {\bf q}, \Omega)$
is given by Eq.(44) with the constraint that the lower cuoff $T$
should be introduced in the frequency integrals, which is due to the
fact that only ${\bf a}_{+}$ fluctuations should be included.
Using the same procedure we used in section IV, we can construct
the equation of motion for the generalized Fermi surface displacement
(in the velocity space)
$u (\theta, {\bf q}, \Omega) = \int {d \omega \over 2 \pi}
\ \delta f (\theta, \omega; {\bf q}, \omega)$.
The corresponding equation is given by Eq.(50) with the change that
$\theta_c$ in the definition of the Landau-interaction-function
$F_{\rm Landau} (\theta)$ is now given by
$\theta_c = {1 \over k_F}
\left ( {\gamma {\rm Max} (\Omega, T) \over \chi}
\right )^{1 \over 1 + \eta}$.
Therefore, the same arguments for the small and large angular momentum
modes can be used to discuss the physical consequences of the QBE and
the change is that the crossover angular momentum is now given by
$l_c \approx 1 / \theta_c \approx
k_F \left ( {\gamma {\rm Max} (\Omega, T) \over \chi}
\right )^{-{1 \over 1 + \eta}}$.

Now let us discuss the effect of the term which depends on
$b_{-} ({\bf r})$. After integration $\int d \omega  / 2 \pi$
of the QBE for the generalized distribution function
$\delta f (\theta, \omega; {\bf r}, t)$,
this term has the following form in the QBE for
$u (\theta, {\bf r}, t)$:
$$
{b_{-} ({\bf r}) \over m} \ {\partial \over \partial \theta} \
u (\theta, {\bf r}, t) \ .
\eqno{({\rm A}.4)}
$$
This term provides the scattering mechanism due to ${\bf a}_{-}$
fluctuations and generates a dispersion of the angle $\theta$.
The transport scattering rate $1 / \tau_{-}$ which is due to
${\bf a}_{-}$ fluctuations can be estimated as follows.
In order to examine $b_{-}$ fluctuations, let us first consider
$$
\eqalign{
\langle b_{-} ({\bf q}) b_{-} (-{\bf q}) \rangle
&= \int^{T}_{0} {d \omega \over 2 \pi} \ [ \ n (\omega) + 1 \ ] \
q^2 \ {\rm Im} \ D_{11} (q, \omega) \cr
&\approx \int^{T}_{0} {d \omega \over 2 \pi} \ {T \over \omega} \
q^2 \ {q \omega / \gamma \over
\omega^2 + (\chi q^{1 + \eta} / \gamma)^2} \cr
&\approx q^3 / \gamma \hskip 0.3cm {\rm for} \ \ q \le q_0 \ ,
}
\eqno{({\rm A}.5)}
$$
where $q_0 = (\gamma T / \chi)^{1 \over 1 + \eta}$.
Therefore, the typical length scale of $b_{-} ({\bf r})$ fluctuations
is given by $l_0 = 1 / q_0$.
The typical value of $b_{-} ({\bf r})$ over the length scale $l_0$
can be estimated from
$\langle b_{-} ({\bf r}) b_{-} ({\bf r'}) \rangle
\approx 1 / (\gamma l^5_0)$
for $|{\bf r} - {\bf r'}| \le l_0$ so that
typical $b_{-} \approx 1 / \sqrt{\gamma l^5_0}$.
The dispersion of the angle $\Delta \theta$ after the fermion
travels over the length $l_0$ can be estimated as
$\Delta \theta = (b_{-} / m) \Delta t \approx
1 / (\sqrt{\gamma l^5_0} \ m) \ (l_0 / v_F) \approx 1 / (k_F l_0)^{3/2}$.
Let $l_{\rm M} = n l_0$ be the mean free path which is defined
by the length scale after which the total dispersion of the
angle becomes of the order one.
The number $n$ can be estimated by requiring that the total
dispersion of the angle
$\sqrt{n} \ \Delta \theta \approx {\sqrt{n} / (k_F l_0)^{3/2}}$
becomes of the order one so that $n \approx (k_F l_0)^{3}$.
Thus, $l_{\rm M} \approx k^3_F l^4_0$.
{}From $l_{\rm M} = v_F \tau_{-}$, the scattering rate due to
${\bf a}_{-}$ fluctuations can be estimated as
$1 / \tau_{-} \propto T^{4 \over 1 + \eta}$.

Note that $1 / \tau_{-} \propto T^{4 \over 1 + \eta}$ is the same
order as that of the scattering rate due to
${\bf a}_{+}$ fluctuations in the
case of the small angular momentum modes ($l < l_c$).
For $l < l_c$, the contribution from the imaginary part of
the self-energy ${\rm Im} \ \Sigma^R \propto T^{2 \over 1 + \eta}$
is canceled by
the contribution from the Landau-interaction function
so that the resulting scattering rate is proportional to
$T^{4 \over 1 + \eta}$.
In the other limit of  large angular momentum modes ($l > l_c$),
$1 / \tau_{-}$
can be completely ignored. This is because
the self-energy contribution dominates.
Since $1 / \tau_{-} < T$ and it is at most the same order as
the scattering rate due to ${\bf a}_{+}$ fluctuations even
in the case of the small angular momentum modes,
ignoring this contribution does not affect the general
consequences of the QBE, which are discussed
in sections IV, V, and VI.

Therefore, the QBE for the generalized distribution function
at finite temperatures is essentially given by
Eq.(44) with the lower cutoff $T$ of the frequency integral
in the expression of the contributions from the self-energy
and the Landau-interaction-function.
As a result, the form of the QBE is the same as that of
the zero temperature  case and the only difference is that
the crossover angle $\theta_c$ and the crossover angular momentum
$l_c$ are now given by
$\theta_c \approx {1 \over k_F} \ \left ( {\gamma \
{\rm Max} (\Omega, T) \over \chi} \right )^{1 \over 1 + \eta}$ and
$l_c \approx 1 / \theta_c \propto
\left [ \ {\rm Max} (\Omega, T) \ \right ]^{-{1 \over 1 + \eta}}$
respectively.

\vfill\vfill\vfill
\break

\vskip 0.5cm

\centerline{\bigbf References}

\vskip 0.5cm

\item{[1]} H. W. Jiang {\it et al.}, Phys. Rev. {\bf B} {\bf 40}, 12013 (1989).
\item{[2]} R. R. Du {\it et al.}, Phys. Rev. Lett. {\bf 70}, 2944 (1993);
D. R. Leadley {\it et al.}, Phys. Rev. Lett. {\bf 72}, 1906 (1994).
\item{[3]} R. R. Du {\it et al.}, Phys. Rev. Lett. {\bf 73}, 3274 (1994);
H. C. Manoharan, M. Shayegan, and S. J. Klepper {\it et al.},
Phys. Rev. Lett. {\bf 73}, 3270 (1994).
\item{[4]} J. K. Jain, Phys. Rev. Lett. {\bf 63}, 199 (1989);
Phys. Rev. {\bf B} {\bf 41}, 7653 (1990); Adv. Phys. {\bf 41}, 105 (1992).
\item{[5]} A. Lopez and E. Fradkin, Phys. Rev. {\bf B} {\bf 44},
5246 (1991); Phys. Rev. Lett. {\bf 69}, 2126 (1992).
\item{[6]} B. I. Halperin, P. A. Lee, and N. Read,
Phys. Rev. {\bf B} {\bf 47}, 7312 (1993).
\item{[7]} V. Kalmeyer and S. C. Zhang, Phys. Rev. {\bf B} {\bf 46},
9889 (1992).
\item{[8]} R. L. Willet {\it et al.}, Phys. Rev. Lett. {\bf 71}, 3846 (1993);
R. L. Willet {\it et al.}, Phys. Rev. {\bf B} {\bf 47}, 7344 (1993).
\item{[9]} W. Kang {\it et al.}, Phys. Rev. Lett. {\bf 71}, 3850 (1993).
\item{[10]} V. J. Goldman, B. Su, and J. K. Jain, Phys. Rev. Lett. {\bf 72},
2065 (1994).
\item{[11]} N. Nagaosa and P. A. Lee, Phys. Rev. Lett. {\bf 64}, 2550 (1990);
P. A. Lee and N. Nagaosa, Phys. Rev. {\bf B} {\bf 46}, 5621 (1992).
\item{[12]} Y. B. Kim, A. Furusaki, X.-G. Wen, and P. A. Lee,
Phys. Rev. {\bf B} {\bf 50}, 17917 (1994).
\item{[13]} B. L. Altshuler, L. B. Ioffe, and A. J. Millis,
Phys. Rev. {\bf B} {\bf 50}, 14048 (1994).
\item{[14]} Y. B. Kim, P. A. Lee, X.-G. Wen, and P. C. E. Stamp,
{\it Influence of gauge field fluctuations on composite fermions
near the half-filled state}, cond-mat/9411057.
\item{[15]} A. Stern and B. I. Halperin, {\it Singularities in the
Fermi liquid description of a partially filled Landau level and
energy gaps of fractional quantum Hall states}, cond-mat/9502032.
\item{[16]} B. L. Altshuler and L. B. Ioffe, Phys. Rev. Lett. {\bf 69},
2979 (1992).
\item{[17]} D. V. Khveshchenko, R. Hlubina, and T. M. Rice,
Phys. Rev. {\bf B} {\bf 48}, 10766 (1993).
\item{[18]} D. V. Khveshchenko and P. C. E. Stamp, Phys. Rev. Lett. {\bf 71},
2118 (1993); Phys. Rev. {\bf B} {\bf 49}, 5227 (1994).
\item{[19]} J. Polchinski, Nucl. Phys. {\bf B 422}, 617 (1994).
\item{[20]} Junwu Gan and Eugene Wong, Phys. Rev. Lett. {\bf 71},
4226 (1994).
\item{[21]} L. B. Ioffe, D. Lidsky, and B. L. Altshuler,
Phys. Rev. Lett. {\bf 73}, 472 (1994).
\item{[22]} H. -J. Kwon, A. Houghton, and J. B. Marston,
Phys. Rev. Lett. {\bf 73}, 284 (1994); Brown University Preprint,
{\it Theory of fermion liquid}, cond-mat/9501067.
\item{[23]} C. Nayak, and F. Wilczek, Nucl. Phys. {\bf B 417}, 359 (1994);
Nucl. Phys. {\bf B 430}, 534 (1994).
\item{[24]} S. He, P. M. Platzman, and B. I. Halperin,
Phys. Rev. Lett. {\bf 71}, 777 (1993);
Y. Hatsugai, P.-A. Bares, and X.-G. Wen,
Phys. Rev. Lett. {\bf 71}, 424 (1993);
Y. B. Kim and X.-G. Wen, Phys. Rev. {\bf B} {\bf 50},
8078 (1994).
\item{[25]} R. E. Prange and L. P. Kadanoff, Phys. Rev. {\bf 134},
A566 (1964).
\item{[26]} L. P. Kadanoff and G. Baym,
{\it Quantum Statistical Mechanics}, Benjamin, New York, 1962.
\item{[27]} L. V. Keldysh, Zh. Eksp. Teor. Fiz. {\bf 47}, 1515 (1964)
[Sov. Phys. - JETP {\bf 20}, 1018 (1965)].
\item{[28]} G. D. Mahan, {\it Many Particle Physics}, 2nd Edition, Plenum,
New York, 1990;
J. Rammer and H. Smith, Rev. Mod. Phys. {\bf 58}, 323 (1986).
\item{[29]} Y. H. Chen, F. Wilczek, E. Witten, and B. I. Halperin,
Int. J. Mod. Phys. {\bf 3}, 1001 (1989).
\item{[30]} S. H. Simon and B. I. Halperin, Phys. Rev. {\bf B} {\bf 48},
17368 (1993); S. H. Simon and B. I. Halperin, Phys. Rev. {\bf B} {\bf 50},
1807 (1994); Song He, S. H. Simon, and B. I. Halperin,
Phys. Rev. {\bf B} {\bf 50}, 1823 (1994).
\item{[31]} W. H\"ansch and G. D. Mahan, Phys. Rev. {\bf B} {\bf 28},
1886 (1983).
\item{[32]} R. L. Willet, K. W. West, and L. N. Pfeiffer, preprint.

\vfill\vfill\vfill
\break

\vskip 0.5cm

\centerline{\bigbf Figure captions}

\vskip 0.5cm

\item{Fig.1}
The one-loop Feynman diagram for the self-energy of the fermions.
Here the solid line represents the fermion propagator and
the wavy line denotes the RPA gauge field propagator.

\vskip 0.5cm

\item{Fig.2}
A broad wave packet (a) and a narrow wave packet (b)
(given by the shaded region) created in the momentum space.
The circle is the schematic representation of the Fermi
surface, which is actually not so well defined, and
the arrow represents the direction of motion of the
wave packet.

\vskip 0.5cm

\item{Fig.3}
The energy band $\Omega (\theta)$ of the tight binding model
given by Eq.(61) as a function of $\theta$.
The shaded region around the center of the band corresponds
to the continuum states and the hatched region in the tails
of the band corresponds to the bound states.

\vskip 0.5cm

\item{Fig.4}
The elementary excitations in $\Omega - q$ space in the absence
of the collision integral.
The shaded region corresponds to the particle-hole continuum
and the hatched region corresponds to the collective modes.
The boundary is given by the singular
dispersion relation $\Omega \propto q^{1 + \eta \over 2}$ for
$1 < \eta \le 2$ and $\Omega \propto q / |{\rm ln} \ q|$ for
$\eta = 1$.

\vskip 0.5cm

\item{Fig.5}
The lowest excitation spectrum of the composite fermion system
in the presence of the finite effective magnetic field $\Delta B$
as a function of the wave vector $q$ (solid line).
The dashed line is the scaling curve described in the text.
For $q \gg q_c$, the excitation gap becomes smaller and
is proportional to
$|\Delta B|^{1 + \eta \over 2}$ for $1 < \eta \le 2$ and
$|\Delta B| / |{\rm ln} \ \Delta B|$ for $\eta = 1$.
$q_c \propto \sqrt{|\Delta B|}$ for $1 < \eta \le 2$ and
$q_c \propto \sqrt{|\Delta B| \ |{\rm ln} \ \Delta B|}$ for
$\eta = 1$.

\bye